\documentclass[fleqn,usenatbib]{mnras}

\usepackage{newtxtext,newtxmath}
\usepackage{adjustbox}
\usepackage{graphics}
\usepackage[T1]{fontenc}
\usepackage{rotating} 
\DeclareRobustCommand{\VAN}[3]{#2}
\let\VANthebibliography\thebibliography
\def\thebibliography{\DeclareRobustCommand{\VAN}[3]{##3}\VANthebibliography}

\usepackage{graphicx}	
\usepackage{amsmath}	
\usepackage{natbib}
\usepackage{pdflscape}

\newcommand\hi{\mbox{H\,{\sc i}}\ }
\newcommand\hix{\mbox{H\,{\sc i}}}

\newcommand\nhi{$N_{\rm HI}$}

\newcommand\prospect{\textsc{ProSpect\ }}

\title[\mbox{H\,{\sc i}} absorption in GAMA galaxies]{FLASH Pilot Survey: Detections of associated 21\,cm \mbox{H\,{\sc i}} absorption in GAMA galaxies at 0.42 $<z<$ 1.00}

\author[R. Su et al.]{
Renzhi Su$^{1,2,3}$, 
Elaine M. Sadler$^{3,4,5}$,\thanks{E-mail: elaine.sadler@sydney.edu.au}
James R. Allison$^{6,5}$,
Elizabeth K. Mahony$^{3}$,
Vanessa A. Moss$^{3}$, 
\newauthor
Matthew T. Whiting$^{3}$, 
Hyein Yoon$^{4,5}$,  
J. N. H. S. Aditya$^{4,5}$,
Sabine Bellstedt$^{7}$,
Aaron S. G. Robotham$^{7,5}$,
\newauthor
Lilian Garratt-Smithson$^{7}$,
Minfeng Gu$^{1}$,
B$\ddot{\rm a}$rbel S. Koribalski$^{3,5,8}$,
Roberto Soria$^{2,4}$,
Simon Weng$^{4,5}$
\newauthor
\\
$^{1}$Key Laboratory for Research in Galaxies and Cosmology, Shanghai Astronomical Observatory, Chinese Academy of Sciences, \\80 Nandan 
Road, Shanghai 200030, China\\
$^{2}$University of Chinese Academy of Sciences, 19A Yuquan Road, Beijing 100049, China \\
$^{3}$ATNF, CSIRO Space and Astronomy, PO Box 76, Epping, NSW 1710, Australia \\
$^{4}$Sydney Institute for Astronomy, School of Physics A28, University of Sydney, NSW 2006, Australia \\
$^{5}$ARC Centre of Excellence for All Sky Astrophysics in 3 Dimensions (ASTRO 3D) \\
$^{6}$First Light Fusion Ltd., Unit 9/10 Oxford Industrial Park, Mead Road, Yarnton, Kidlington OX5 1QU, UK\\
$^{7}$International Centre for Radio Astronomy Research (ICRAR), University of Western Australia, Crawley, WA 6009, Australia\\
$^{8}$ School of Science, Western Sydney University, Locked Bag 1797, Penrith, NSW 2751, Australia \\
}
\date{Accepted XXX. Received YYY; in original form ZZZ}

\pubyear{2021}

\begin{document}
\label{firstpage}
\pagerange{\pageref{firstpage}--\pageref{lastpage}}
\maketitle

\begin{abstract}
We present the results of a search for associated 21\,cm \hi absorption at redshift $0.42<z<1.00$ in radio-loud galaxies from three Galaxy And Mass Assembly (GAMA) survey fields.  These observations were carried out as part of a pilot survey for the ASKAP First Large Absorption Survey in \hi (FLASH). 
From a sample of 326 radio sources with 855.5\,MHz peak flux density above 10\,mJy we detected two associated \mbox{H\,{\sc i}} absorption systems, in SDSS\,J090331+010847 at $z=0.522$ and SDSS\,J113622+004852 at $z=0.563$.  
Both galaxies are massive (stellar mass $>10^{11}$M$_\odot$) and have optical spectra characteristic of luminous red galaxies, though SED fitting implies that SDSS\,J113622+004852 contains a dust-obscured starburst with SFR $\sim69$\,M$_\odot$\,yr$^{-1}$. 
The \hi absorption lines have a high optical depth, with $\tau_{\rm pk}$ of $1.77\pm0.16$ for SDSS\,J090331+010847 (the highest value for any $z>0.1$ associated system found to date) and $0.14\pm0.01$ for SDSS\,J113622+004852. 
In the redshift range probed by our ASKAP observations, the detection rate for associated \mbox{H\,{\sc i}} absorption lines (with $\tau_{\rm pk}>0.1$ and at least 3$\sigma$ significance) is $2.9_{-2.6}^{+9.7}$ percent. Although the current sample is small, this rate is consistent with a trend seen in other studies for a lower detection rate of associated 21\,cm \hi absorption systems at higher redshift. 
We also searched for OH absorption lines at  $0.67<z<1.34$, but no detection was made in the 145 radio sources searched. 
\end{abstract}



\maketitle

\section{Introduction}
Hydrogen is the most common element in the Universe and neutral hydrogen provides the main fuel for star formation in galaxies, which is known to evolve across cosmic time (e.g. \citealt{hopkins2006,driver2018}). In the central parts of galaxies, infalling gas can fuel Active Galactic Nuclei (AGN) (e.g. \citealt{dressel1983,taylor1999,araya2010,maccagni2014}) and transfer the energy released by AGN to the surrounding interstellar medium (ISM) in host galaxies, a process known as feedback (e.g. \citealt{morganti2003,morganti2005, morganti2013,kanekar2008,rodriguez2009,teng2013}) which can strongly influence the subsequent evolution of the galaxy. The distribution, mass and kinematics of neutral hydrogen in galaxies can therefore advance our understanding of their assembly and evolution.

In its neutral atomic state, hydrogen can be traced in the radio band (either in emission or in absorption) through a hyperfine transition at a rest-frame frequency of 1420.4\,MHz. 21 cm line emission can provide information, e.g. \mbox{H\,{\sc i}} mass, morphology, and dynamics, for galaxies, but its faintness, e.g. compared to CO emission, makes the relevant studies difficult beyond the local Universe ($z<0.1$) \citep[e.g.][]{lang2003,chung2009,lee2012,van2016,winkel2016,yoon2017,catinella2018,koribalski2020}. 

In contrast to \mbox{H\,{\sc i}} emission, the detection sensitivity for \mbox{H\,{\sc i}} absorption is independent of redshift and depends only on the brightness of the background radio source used as a probe \citep{morganti2018}. The disadvantage, however, is that only the neutral hydrogen along our sight to the radio continuum source can be traced. \hi absorption searches can detect both intervening lines (which trace neutral gas in and around the general population of distant galaxies) and associated lines (which allow us to trace the kinematics of neutral gas within radio-loud AGN). 

Associated \mbox{H\,{\sc i}} absorption searches have already been carried out for diverse samples of objects, including compact radio sources (i.e. compact steep-spectrum (CSS) sources and gigahertz-peaked spectrum (GPS) sources), extended sources (Fanaroff-Riley class I (FR I) and class II (FR II) radio galaxies), Seyferts and dust obscured quasars 
(\citealt{baan1981,mirabel1989,van1989,vermeulen2003,curran2013a,gupta2006,struve2010,allison2012,allison2014,allison2016,maccagni2017,allison2020,koribalski2012,gereb2015,aditya2018b,glowacki2017,glowacki2019,dutta2019,murthy2021,murthy2022}). A useful review of the results  is given by \cite{morganti2018}. 

Most of the associated \mbox{H\,{\sc i}} absorption detections made so far are at redshift $z<1$, and some detailed results have been achieved from very long baseline interferometry (VLBI) observations, such as the finding of circumnuclear disks (e.g. \citealt{taylor1996,van2000,struve2010}), the identification of AGN accretion (e.g. \citealt{maccagni2014,taylor1999,araya2010}), and characterising the jet-gas interactions (e.g. \citealt{morganti2003,kanekar2008,rodriguez2009,mahony2013,schulz2021}). The gas structures that can absorb background radio continuum are diverse, and in addition to circumnuclear disks, outflowing and infalling gas, the associated \mbox{H\,{\sc i}} absorption may also arise from the galactic disk (e.g. \citealt{morganti2002}) or from high speed clouds that move across our sight to radio continuum background (e.g. \citealt{conway1999}). These \mbox{H\,{\sc i}} absorption detections have advanced our understanding of star formation, AGN structures and co-evolution between SMBH and host galaxy.

Associated \mbox{H\,{\sc i}} absorption studies at higher redshift ($z>1$) have so far yielded only a few detections despite significant effort (e.g. \citealt{curran2008,curran2013b,curran2016,aditya2018b,grasha2019} and references therein).  To date, only 12 associated \mbox{H\,{\sc i}} absorption systems have been detected at $z>1$ \citealt{uson1991,moore1999,ishwara2003,curran2013b, aditya2017,aditya2018a,Chowdhury2020,Dutta2020,aditya2021,gupta2021}), and the detection rate here is much lower than in the local Universe. \cite{curran2008} pointed out that the sources selected at high redshift generally have high ultraviolet (UV) luminosity ($L_{\rm UV}$ > $10^{23}$\,W\,Hz$^{-1}$) which can, in principle, ionise all the hydrogen gas in host galaxies and hence results in the low detection rate for associated absorption at high redshift (e.g. \citealt{curran2008,curran2012}). Some intervening \mbox{H\,{\sc i}} absorption systems have also been detected at $z>1$ (e.g. \citealt{moore1999,srianand2008,kanekar2009,gupta2012}), but most studies of neutral hydrogen in the distant universe have used Ly $\alpha$ absorption from optical observations (e.g. \citealt{noterdaeme2012,lee2013,crighton2015}).

The First Large Absorption Survey in HI \citep[FLASH;][]{allison2022} is a wide-field survey for \mbox{H\,{\sc i}} absorption that will cover the entire southern sky in the redshift range $0.42<z<1.00$, enabled by the Australian SKA Pathfinder (ASKAP) advantages of wide field of view and 288\,MHz instantaneous bandwidth. Before the full operation of the 36-antenna ASKAP, early FLASH results using a sub-array of ASKAP dishes have demonstrated the capability of this telescope for \hi absorption studies (\citealt{allison2015,allison2017,moss2017,allison2020,sadler2020,mahony2021}). 
In this paper, we present the results of a search for associated  \mbox{H\,{\sc i}} absorption at $0.42<z<1.00$ in the three equatorial GAMA fields (GAMA 09, GAMA 12, and GAMA 15), using data from the pilot FLASH survey. Throughout the paper, we adopt a $\Lambda$CDM cosmology with $\rm H_{0}=70\,km\,s^{-1}\,Mpc^{-1}$,  $\Omega_{m}$=0.3, and $\Omega_{\Lambda}$=0.7.
 
\section{The GAMA survey fields}

\subsection{The GAMA survey}
The three equatorial fields (G09, G12 and G15) from the GAMA Galaxy Survey \citep{driver2011,liske2015} were chosen for the FLASH pilot study because they overlap with the Sloan Digital Sky Survey (SDSS\footnote{\url{https://www.sdss.org}}) but also  provide deeper spectroscopic redshift coverage and additional multi-wavelength data. 

\begin{table*}
\centering
\begin{tabular}{llllll}
\hline
Field & \multicolumn{1}{c}{RA[deg]} & \multicolumn{1}{c}{Dec[deg]} & \multicolumn{1}{c}{RA[hms]} & \multicolumn{1}{c}{Dec[dms]} & Area  \\
   & \multicolumn{2}{c}{(J2000)} & \multicolumn{2}{c}{(J2000)} & deg$^2$  \\
\hline
\multicolumn{4}{l}{\it Optical GAMA spectroscopic fields} \\
  G09  & 129.0 to 141.0 & $-2.0$ to $+3.0$ & 08 36 00.0 to 09 24 00.0 & $-02$ 00 00 to $+03$ 00 00 & 60 \\
  G12  & 174.0 to 186.0 & $-3.0$ to $+2.0$ & 11 36 00.0 to 12 24 00.0 & $-03$ 00 00 to $+02$ 00 00 & 60 \\
  G15  & 211.5 to 223.5 & $-2.0$ to $+3.0$ & 14 06 00.0 to 14 54 00.0 & $-02$ 00 00 to $+03$ 00 00 & 60 \\
 \multicolumn{4}{l}{\it ASKAP FLASH pilot fields} \\
  G09A  & 128.7 to 135.1 & $-2.7$ to $+3.7$ & 08 34 47.6 to 09 00 23.6 & $-02$ 42 00 to $+03$ 42 00 & 40 \\
  G09B  & 134.9 to 141.3 & $-2.7$ to $+3.7$ & 08 59 54.1 to 09 25 13.2 & $-02$ 42 00 to $+03$ 42 00 & 40 \\
  G12A & 173.7 to 180.1 & $-3.7$ to $+2.7$ & 11 34 47.1 to 12 00 23.2 & $-03$ 42 00 to $+02$ 42 00 & 40 \\
  G12B & 179.9 to 186.3 & $-3.7$ to $+2.7$ & 11 59 36.8 to 12 25 12.8 &  $-03$ 42 00 to $+02$ 42 00 & 40 \\
  G15A & 210.9 to 217.3 & $-2.7$ to $+3.7$ & 14 03 45.1 to 14 29 21.1 & $-02$ 42 00 to $+03$ 42 00 & 40 \\
  G15B & 217.1 to 223.5 & $-2.7$ to $+3.7$ & 14 28 24.0 to 14 54 00.0 & $-02$ 42 00 to $+03$ 42 00 & 40 \\
\hline
\end{tabular}
\caption{Location of the GAMA spectroscopic fields and ASKAP pilot survey fields used for this study.}
\label{tab:gama_range}
\end{table*}

The GAMA spectroscopic survey was carried out in two phases (GAMA I, II) over the period 2008--2014, with targeted objects selected from various catalogues including SDSS DR 6, 7, 8 and the UKIRT Infrared Deep Sky Survey (UKIDSS)  \citep{liske2015}. The first phase (GAMA I) surveyed objects in the G09, G12 and G15 fields down to magnitude limits of $r < 19.4$ mag in G09 and G15, and $r < 19.8$ mag in G12. The second phase (GAMA II) increased the original areas of G09, G12 and G15 to cover a total  area of 180\,deg$^2$ as listed in Table \ref{tab:gama_range},  and included all galaxies down to a magnitude limits of $r < 19.8$. Two southern fields, G02 and G23, were added in GAMA II. 

The GAMA spectra cover a wavelength range of $3700\sim8800$ {\AA} with a resolution of 4.6\,{\AA}. Redshifts were measured by the GAMA team using the codes $RUNZ$ \citep{driver2011} and $AUTOZ$ \citep{baldry2014}, and a value $Q$ was assigned to quantify the reliability of each redshift. For our purposes, a value of $Q\geq3$ means that the redshift is reliable and can be used in scientific analyses, while $Q<3$ means that the redshift may be unreliable. 

\subsection{Radio continuum studies in the GAMA fields}
There have been at least two previous studies of radio continuum sources in the GAMA equatorial fields. 
\cite{prescott2016} carried out a deep radio survey at 325\,MHz with the Giant Metrewave Radio Telescope (GMRT), covering 138\,deg$^2$ of sky in the G09, G12 and G15 GAMA fields. They matched their radio data with optical imaging and redshifts from the GAMA survey, and compiled a catalogue of 499 GAMA/GMRT matches out to redshift $z=0.6$ in the GAMA fields. 

Secondly, \cite{ching2017} published the Large Area Radio Galaxy Evolution Spectroscopic Survey (LARGESS), which covered $\sim800$ deg$^{2}$ of the sky and provided a spectroscopic and photometric catalogue of radio-source hosts. The catalogue was based on cross-matching SDSS and FIRST data and included both galaxies and QSOs out to redshift $z\sim0.8$ and in some cases beyond. The full LARGESS catalogue contains spectroscopic data for 12,329 radio sources and includes SDSS IDs, coordinates, optical magnitudes, 1.4\,GHz radio flux density, redshift, and spectral-line measurements. All three GAMA equatorial fields (G09, G12, and G15) were included, and \cite{ching2017} obtained additional optical spectra for radio-source hosts that were either stellar in optical images (candidate QSOs) or galaxies fainter than the GAMA magnitude limit. The LARGESS catalogue includes 3668 matched radio sources within the GAMA fields listed in Table \ref{tab:gama_range}, and a spectroscopic completeness of 86--91 per cent was achieved for GAMA objects down to SDSS $i$ mag = 20.5 in these three fields. \cite{ching2017} also assigned a $Q$ value to each source, using the same convention as the GAMA team, to indicate the reliability of its redshift measurement.

We took the \cite{ching2017} catalogue as a starting point for our study, but also used radio continuum data from our own FLASH observations to carry out additional radio-optical cross-matching as described in the next section.

\section{ASKAP observations}

\subsection{The ASKAP telescope}
The ASKAP telescope \citep{deboer2009, hotan2021} is located at the radio-quiet Murchison Radio-astronomy Observatory (MRO) in Western Australia. It consists of an array of 36 antennas, each 12\,m in diameter, with a maximum baseline of 6\,km and a total collecting area of $\sim4000$ m$^{2}$. ASKAP has an instantaneous bandwidth of 288 MHz, which can be split into channels of 18.5 KHz for spectral-line  observations. A prominent advantage of ASKAP is the installation of phased array feed (PAF) receivers that create up to 36 simultaneous beams and provide a field of view (FOV) of $\sim30$\, deg$^{2}$ (e.g. \citealt{chippendale2015,mcconnell2016,hotan2021}). This combination of large bandwidth and wide FOV makes ASKAP an efficient survey instrument  (\citealt{johnston2007,johnston2008}).

\subsection{Pilot survey observations}
In 2019--20, the FLASH team carried out a 100-hour pilot survey with the full 36 antenna ASKAP array. This time was split into 32 $\times$ 2\,hr pointings covering 980 deg$^{2}$ (6 pointings in the GAMA fields and the remainder other parts of the sky), along with a further 6 $\times$ 6\,hr deep pointings covering the GAMA fields. We here define the fields covered by the 6 pointings as 6 GAMA sub-fields. The aim of the deeper 6\,hr repeat observations was to test whether significant numbers of absorption lines towards faint radio continuum would be missed in the 2\,hr observations planned for the full FLASH survey \citep{allison2022}. 

We observed the 6 GAMA sub-fields with the full 36 antenna  ASKAP array in early 2020, and Table \ref{table:obs_parameters} lists the observing parameters used.  
Each 60\,deg$^2$ GAMA field was covered by two separate ASKAP pointings (A and B) as shown in Table \ref{tab:gama_range},  and Table \ref{table:obs_summary} provides a log of observations. 

\begin{figure*}
    \includegraphics[width=14.8cm]{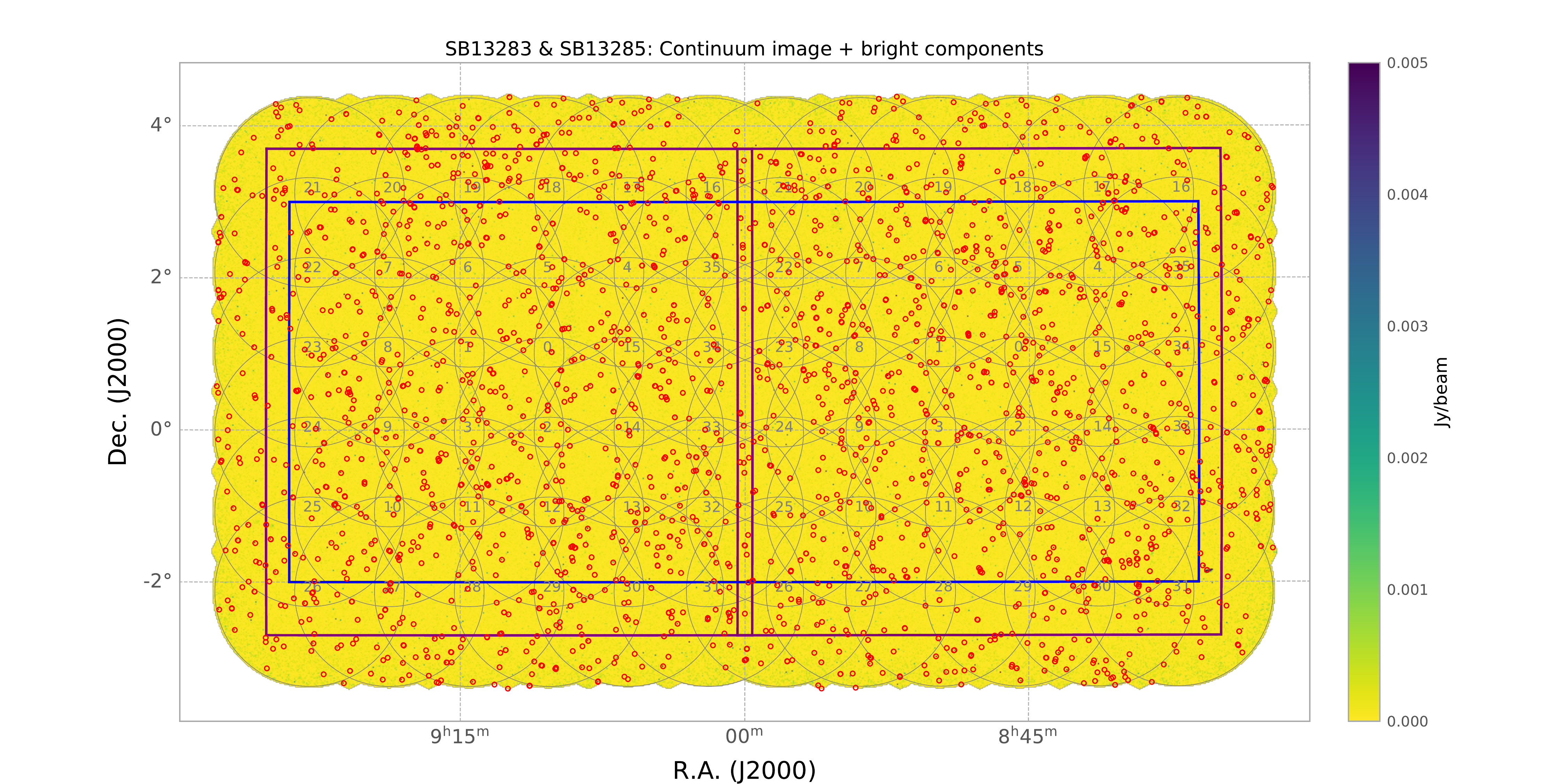}
    \includegraphics[width=14.8cm]{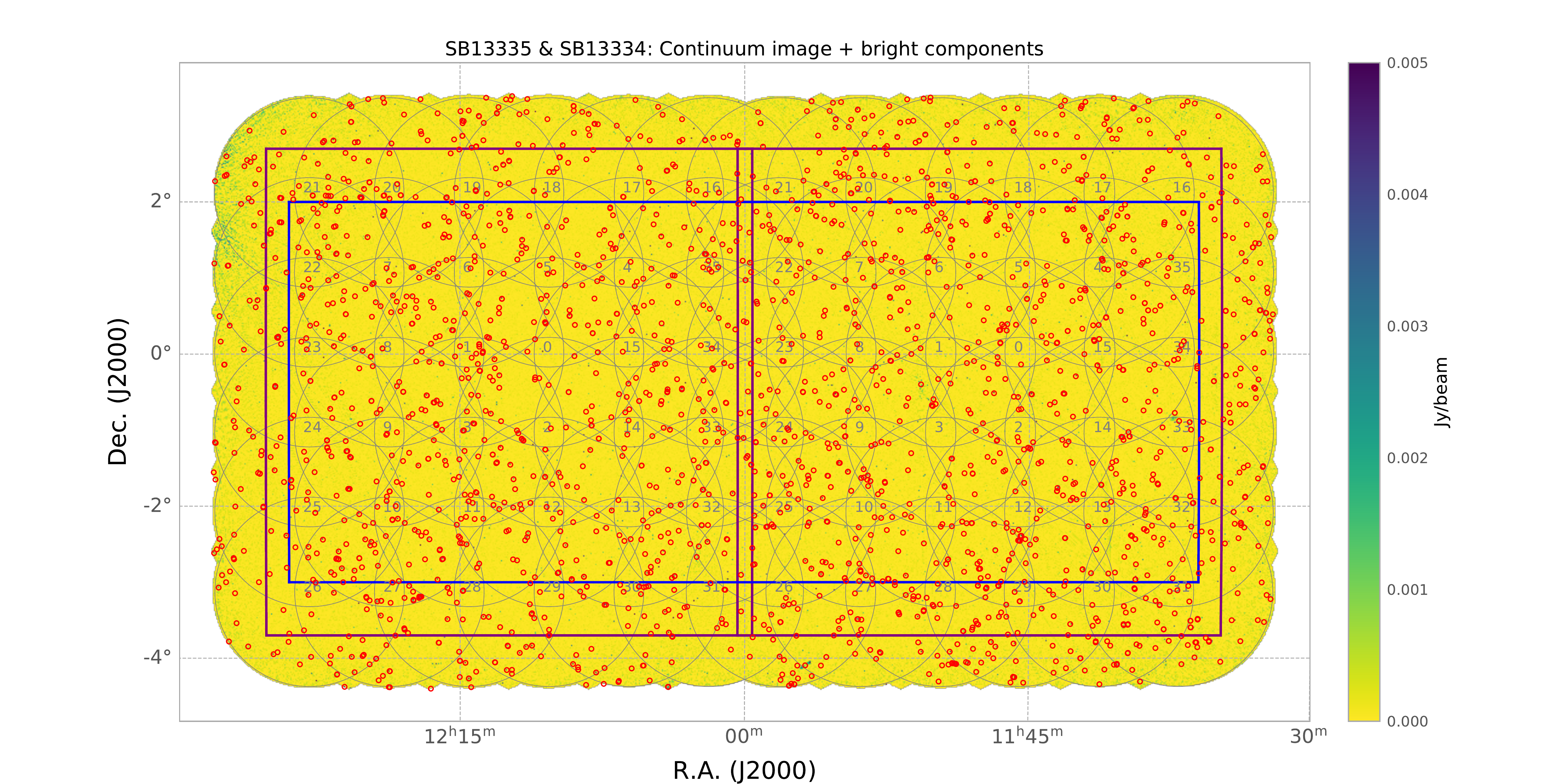}
    \includegraphics[width=14.8cm]{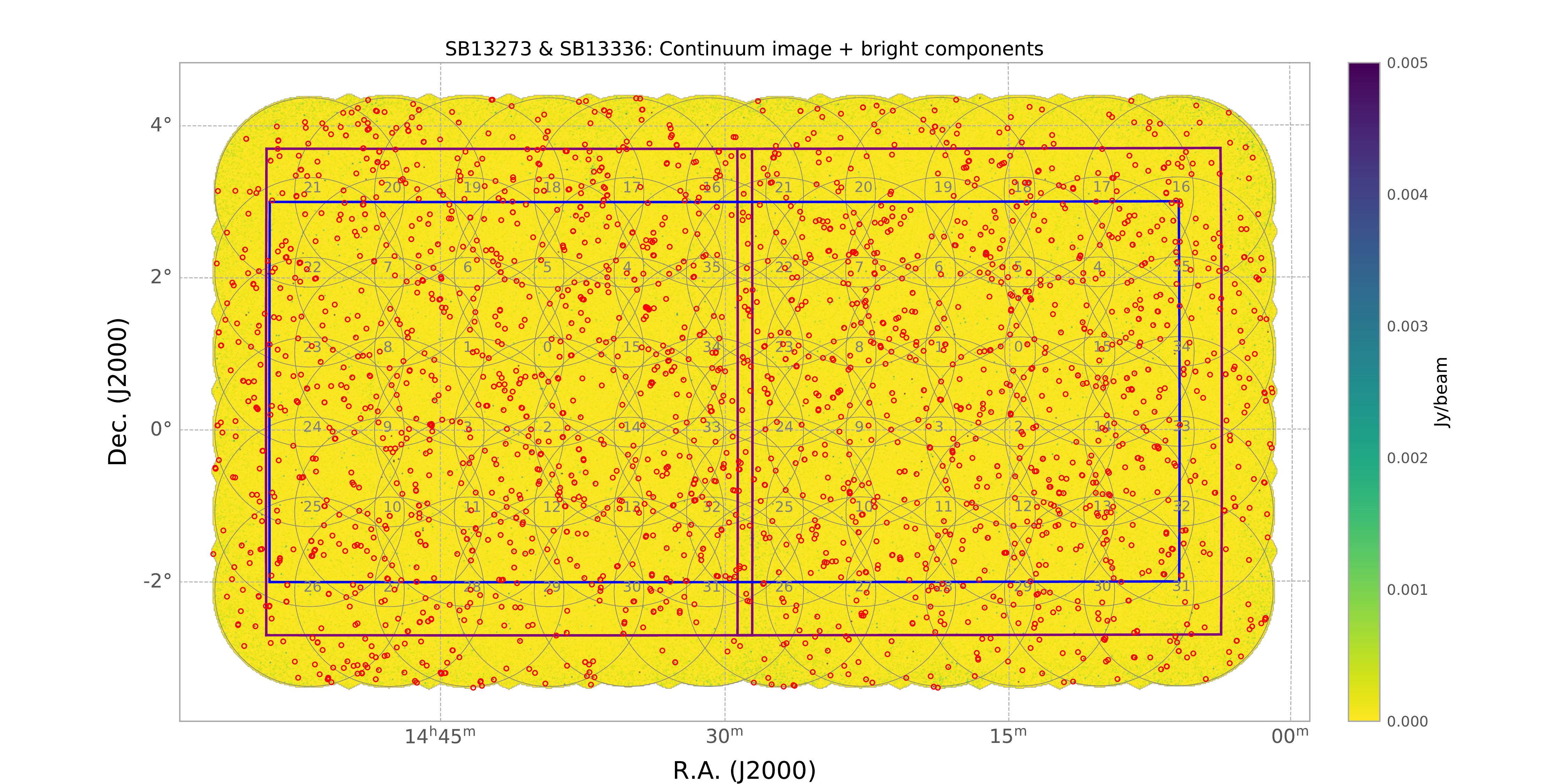}
    \caption{The footprints of 6 GAMA sub-fields. From top to bottom, left to right, they are GAMA\_09B (SBID 13283),  GAMA\_09A (SBID 13285), GAMA\_12B (SBID 13335), GAMA\_12A (SBID 13334), GAMA\_15B (SBID 13273), and GAMA\_15A (SBID 13336).  Red circles show the positions of radio components with peak flux density above 10\,mJy. The purple squares show the 6.4 deg $\times$ 6.4 deg regions in which the rms noise is roughly uniform (Yoon et al., in prep, see also Fig.\ref{fig:spectral_noise}). The 36 beams are plotted with their orders indicated by digits in their centres. The blue rectangles show the optical GAMA G09, G12, and G15 fields. There is a small overlap region between each pair of ASKAP fields, but in general the six fields can be treated independently. }
    \label{fig:GAMA_footprint}
\end{figure*}

We show the footprints of the observations towards these six GAMA sub-fields in Fig. \ref{fig:GAMA_footprint}, where individual radio continuum sources with peak flux density larger than 10\,mJy are also marked. The total area of the FLASH GAMA sub-fields is slightly larger than that of the optical GAMA 09, 12 and 15 fields and (as Fig. \ref{fig:GAMA_footprint} shows) our ASKAP observations fully cover these fields. 

\begin{table}
\begin{tabular}{@{}llcc@{}}
\hline
Central frequency & 855.5\,MHz \\
Instantaneous bandwidth & 288\,MHz \\
Frequency range & 711.5 -- 999.5\,MHz \\
Redshift range for \hi & $z=0.42-1.00$ \\
Redshift range for OH & $z=0.67-1.34$ \\
Spectral channels & 15552 \\
Channel width & 18.5\,kHz \\
Velocity per spectral channel & 5.5 -- 7.8\,km\,s$^{-1}$ \\
rms noise per spectral channel & 3.2 -- 5.1\,mJy\,beam$^{-1}$ \\
\hline
\end{tabular}
\caption{ASKAP observing parameters used for the GAMA fields}
\label{table:obs_parameters}
\end{table}

\begin{table*}
  \caption{Summary of observations of the six GAMA sub-fields. Column [1] is the field name, and Column [2] is the unique CASDA scheduling block identification (SBID) assigned to each observation. Columns [3] and [4] are the RA and Dec. of the pointing centres. Column [5] is the date of observation. Column [6] is the duration of observations (in hours). Column [7] is the width of beamforming used in observations. Columns  [8] and [9] are the the resolution of the continuum image and spectral-line cube respectively. Column [10] is the spectral rms noise per 18.5\,kHz spectral channel in the central part of the frequency band. \\  Two of the six-hour observations (G\_12B and G\_15B) failed at the processing stage because of corrupted data,  
  and these data were not released in CASDA. The six-hour data for G\_09B were processed and released in CASDA, but were not used for an \hi absorption search because of technical issues at the post-processing stage (see \S3.3 of the text).}\label{table:obs_summary}
\begin{tabular}{lcccccccccl}
\hline
Name & SBID & \multicolumn{2}{c}{Pointing Centre} & Obs. date & $t_{\rm obs}$ & BF  & Cont. resolution & Line resolution & $\sigma_{\rm chan}$ \\
& & RA\,[J2000] & Dec.\,[J2000] &  [UTC] & [h] & [MHz] & \multicolumn{2}{c}{[arcsec $\times$ arcsec]} & [mJy\,beam$^{-1}$] \\ 
\relax
[1] & [2] & [3] & [4] & [5] & [6] & [7] & [8] & [9] & [10] \\
\hline
\multicolumn{8}{l}{Two-hour observations}  \\
G\_09A & 13285 & 08:47:35.586&  +00:30:00.00  & 2020\,April\,18 & 2   & 9 & $18.5\times11.7$ & $30\times30$ & 4.96  \\
G\_09B & 13283 & 09:12:25.241& +00:30:00.00 & 2020\,April\,18  & 2  & 9 & $16.7\times13.0$ & $30\times30$  & 4.63  \\
G\_12A & 13334 & 11:47:35.172& -00:30:00.00 & 2020\,April\,21 & 2  & 9  & $16.3\times13.0$ & $30\times30$ & 4.75 &  \\
G\_12B & 13335 & 12:12:24.827& -00:30:00.00 & 2020\,April\,21  & 2 & 9 & $14.3\times13.1$ & $30\times30$ & 4.78 \\
G\_15A & 13336 & 14:16:33.103& +00:30:00.00 & 2020\,April\,21 & 2  & 9 & $14.3\times13.3$ & $30\times30$ & 4.81 \\
G\_15B & 13273 & 14:41:22.758& +00:30:00.00 & 2020\,April\,17 & 2  & 9 & $14.3\times13.3$ & $30\times30$  & 4.83 \\
\multicolumn{8}{l}{Six-hour observations}  \\
G\_09A\_long & 13293 & 08:47:35.586& +00:30:00.00 & 2020\,April\,19 & 6  & 9 & $14.6\times12.9$ & $30 \times30$ & 2.75  \\
G\_09B\_long & 11068 & 09:12:25.241& +00:30:00.00 & 2020\,January\,07 & 6  & 5 & $18.8\times14.0$ & $27.5\times24.8$ & 2.98 & \\
G\_12A\_long & 13306 & 11:47:35.172& -00:30:00.00 & 2020\,April\,20 & 6  & 9 & $15.4\times12.9$ & $30 \times30$ & 2.78 \\
G\_12B\_long & 13366 & 12:12:24.827& -00:30:00.00 & 2020\,April\,22 & 6  & 9 & ... & ...  & ... \\
G\_15A\_long &  13294 & 14:16:33.103& +00:30:00.00 & 2020\,April\,19 & 6  & 9 & $15.1\times12.6$ & $30 \times30$  & 2.83   \\
G\_15B\_long & 13286 & 14:41:22.758& +00:30:00.00 & 2020\,April\,18 & 6  & 9 & .. & ... & ...  \\
\hline
  \end{tabular}
\end{table*}

Our observations used standard ASKAP techniques for observing and calibration, as described by \cite{hotan2021} and \cite{mcconnell2020}. The primary calibration of the array was based on observations of the bright radio source PKS\, B1934-638, which was also used to set the flux density scale \citep{reynolds1994}. 
The spatial resolution set by the longest (6\,km) ASKAP baseline is around 12\,arcsec at our central frequency of 855.5\,MHz, but the longest baselines could not be used in making the spectral-line cubes because of processing constraints. Table \ref{table:obs_summary} lists the final resolution for each observation.  

\subsection{Data processing}
We used the standard ASKAP data reduction pipeline \citep{whiting2020,hotan2021} for the initial processing of our GAMA data. The pipeline reduction includes calibration, continuum and spectral-line imaging, continuum subtraction, production of `component' and `island' continuum source catalogues, and measurement of a range of quality-control parameters. `Islands' are associations of one or more Gaussian components belonging to the same source. as described by \cite{whiting2012}. Fig. \ref{fig:representative_sources} shows some examples of single-component and multi-component sources. 
For multi-component sources, the island position listed by the pipeline (corresponding to the radio centroid for each island) is used to search for the optical host galaxy. A full description of data processing for the FLASH Pilot Survey will be given in a forthcoming data release paper by Yoon et al.\ (in preparation). 

It was discovered during the initial processing that the beam-forming intervals used for FLASH observations (5 or 9\,MHz, see Table \ref{table:obs_summary}) were not implemented as expected and the data showed jumps in phase and amplitude at 1\,MHz intervals. These jumps were corrected in a further stage of data processing by smoothing the bandpass solutions on 1\,MHz intervals and removing any residual artefacts by using 1\,MHz intervals (rather the planned 5 or 9 MHz intervals) for the image-based continuum subtraction. 
At this point, the reduced pipeline data products were released through the CSIRO ASKAP Science Data Archive (CASDA). 

\subsection{Post-processing of the CASDA data}
The FLASH spectral-line data released in CASDA were processed using 1\,MHz intervals when smoothing the bandpass solutions and for image-based continuum subtraction. 
As a result, broad absorption lines with velocity widths above 300\,km\,s$^{-1}$ will be subtracted out by the pipeline processing and not detectable in the spectra provided in CASDA. High S/N lines of any width in the processed spectra may also be distorted by the continuum subtraction procedure of the pipeline. 

To overcome this problem, some further post-processing of the FLASH Pilot Survey data was carried out by the FLASH team as described by Yoon et al.\ (in preparation). We first extracted spectra towards radio sources from initial cube with visibility-based continuum subtraction. Then we fitted the spectra with a polynomial to further remove the continuum over a wider (5 or 9\,MHz) frequency band. These post-processed spectra were used for the current study and will be made available through CASDA in future as part of a wider `value-added' FLASH pilot data release. 

Ongoing technical efforts by the ASKAP team have made significant improvements in observing with larger beam-forming intervals, so it is unlikely that post-processing of this kind will be need for future FLASH observations carried out from mid-2021 onwards.

\section{The GAMA radio-source sample} \label{target_radio_sources_sample}
\subsection{Sample selection}
We now select a well-defined sample of radio galaxies in the GAMA fields that are both sufficiently bright (flux density $>10$\,mJy) and in the right redshift range ($0.42<z<1.00)$ for associated HI absorption to be detectable in our FLASH pilot observations. Our selection criteria are broadly similar to those used in the lower-redshift \hi absorption search by \cite{maccagni2017}. 

\begin{figure}
	\includegraphics[width=\columnwidth]{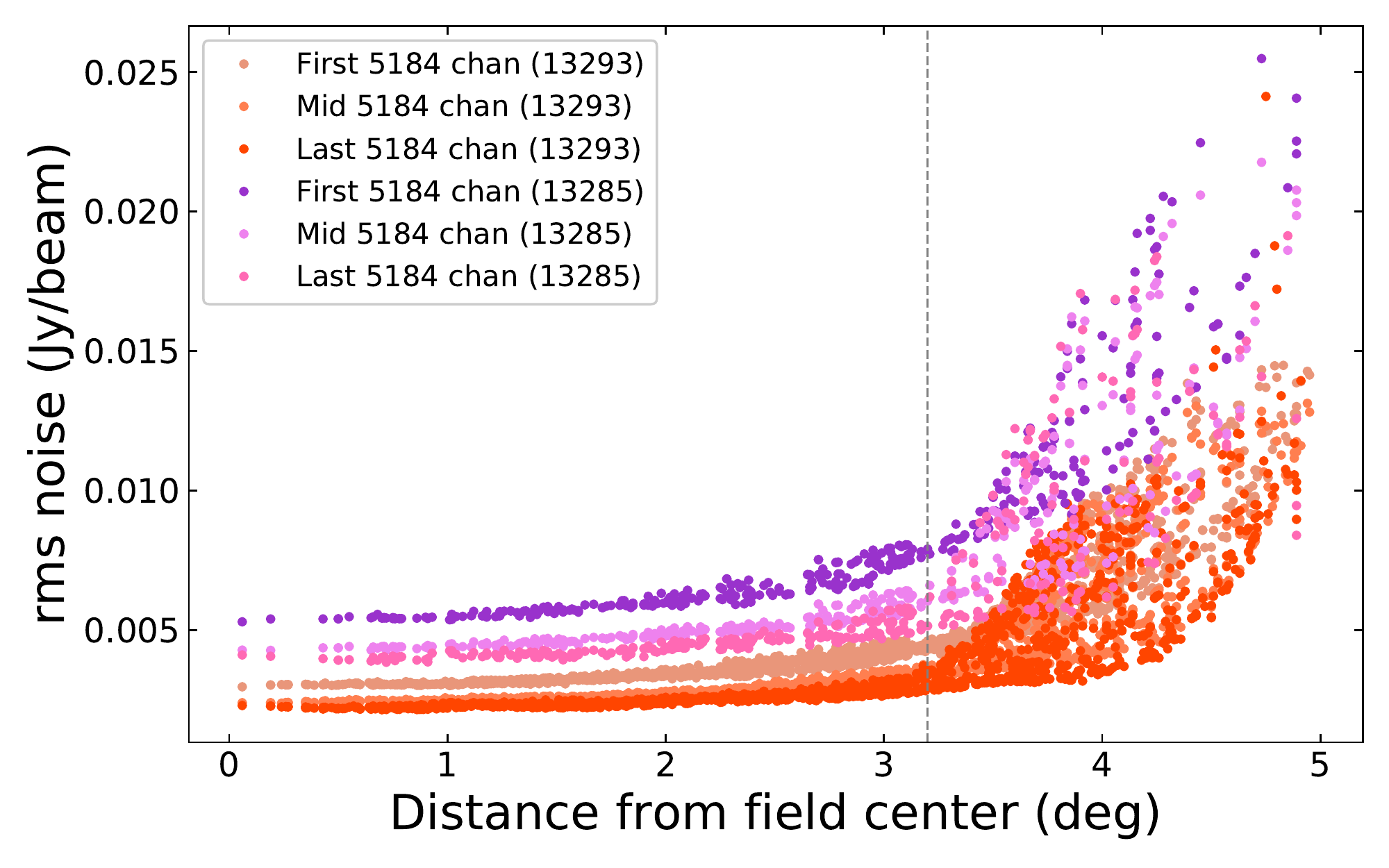}
    \caption{An example of spectral noise distribution as a function of distance from the pointing centre. The data showed are median noise of first 5184 channels, middle 5184 channels, and last 5184 channels of 2 hours observation with SBID 13285 and 6 hours observation with SBID 13293. The vertical grey dashed line marks a distance of 3.2 deg from the pointing centre. It is clear that the spectral noise rises rapidly beyond 3.2 deg. We  show here one 2 hours observation and one 6 hours observation for reference.}
    \label{fig:spectral_noise}
\end{figure}

\begin{figure*}
	\includegraphics[width=18cm]{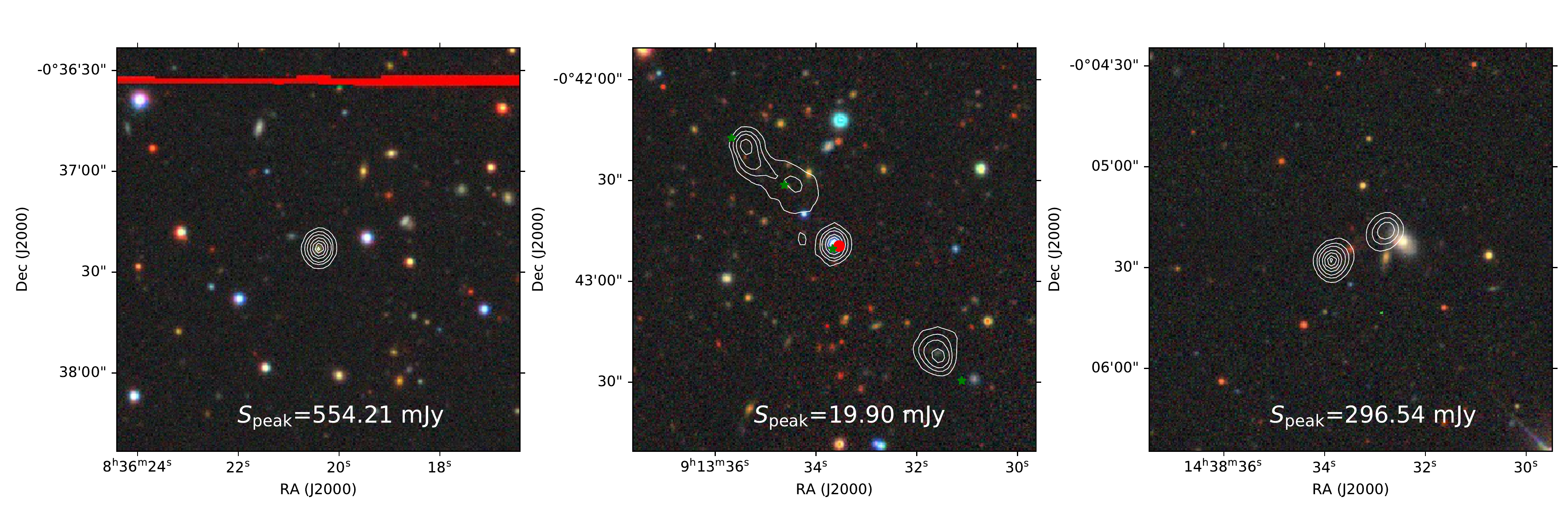}
    \caption{Left: a typical `compact' source with a single continuum component. Middle: an `extended' source with multiple components that was considered suitable for an associated \mbox{H\,{\sc i}} absorption search because of the presence of a bright central core. In our observation, the radio emission was fitted with 4 components with  $S_{855.5} \ge 10$ mJy, indicated by the green stars, when the position of island is marked with the red circle.}   Right: a two-component `extended' source that was not included in our \hi absorption search because of the lack of a central core. In each case, the backgrounds are optical $grz$ images from DESI Legacy Imaging Surveys \citep{dey2019}. All the radio contours plotted are from the 1.4\,GHz FIRST survey, with levels set at at 3, 10, 20, 40, 60, 80, and 95 percent of the FIRST peak flux density.
    \label{fig:representative_sources}
\end{figure*}

The sample was selected as follows: 
\begin{enumerate}
    \item 
We included all sources in the 6.4\,deg $\times$ 6.4\,deg square centred on each pointing centre, i.e. within the purple squares shown in Fig. \ref{fig:GAMA_footprint}, because the spectral-line sensitivity is roughly uniform within this region (Fig. \ref{fig:spectral_noise} shows that the rms noise rises rapidly beyond the boundaries of these squares). 
\item
We then cross-matched the CASDA island catalogue for each field with the SDSS photometric catalogue \citep{alam2015} to identify sources with optical counterparts. For most islands, there are at most a single SDSS counterpart within a 5.0 arcsec radius. For the islands with two or more possible SDSS counterparts, we carefully inspected their corresponding FIRST radio morphology to confirm any association. We then looked for redshift measurements from SDSS, GAMA and NASA/IPAC Extragalactic Database (NED\footnote{\url{http://ned.ipac.caltech.edu/}}) and selected sources with redshifts in the range  $0.42<z<1.00$.  
\item
We then used the \cite{ching2017} catalogue to search for any additional sources that have a separation between islands and optical counterparts greater than 5.0 arcsec or additional redshift measurements that were missed in our second step.
\item
We excluded sources where the only available redshift measurements were flagged as unreliable (i.e. Q value $\leq2$ in either \cite{ching2017} or the GAMA survey).
\item
Finally, we visually inspected all these sources in FIRST images and classified them as either `compact' or  `extended' (including multi-component sources). The  `extended' sources were considered as suitable for associated \mbox{H\,{\sc i}} absorption searching if there was a radio component at the position of the optical galaxy, otherwise they were excluded. Fig.  \ref{fig:representative_sources} shows some representative examples of `compact' and `extended' sources from our final sample.  
\end{enumerate}

This process yielded a final sample of 326 radio sources with optical counterparts in the redshift range $0.42<1.00$ and continuum peak flux densities, $S_{855.5}$, in FLASH observations above 10\,mJy\,beam$^{-1}$. These objects are listed in Table  \ref{table:target-sources} and \ref{table:target-sources-infor}. Table \ref{table:excluded-target-sources} in Appendix A lists the 73 extended sources that were excluded from our \hi absorption search based on their radio properties. For optical objects with multiple radio components, we used the ASKAP spectrum from the component that is closest to optical coordinates to search for \hi absorption. For the fields where data from both 2 hour and 6 hour observations are available, we used the spectrum from the deeper 6 hour observation in our initial search. 

\begin{figure*}
    \includegraphics[width=14cm]{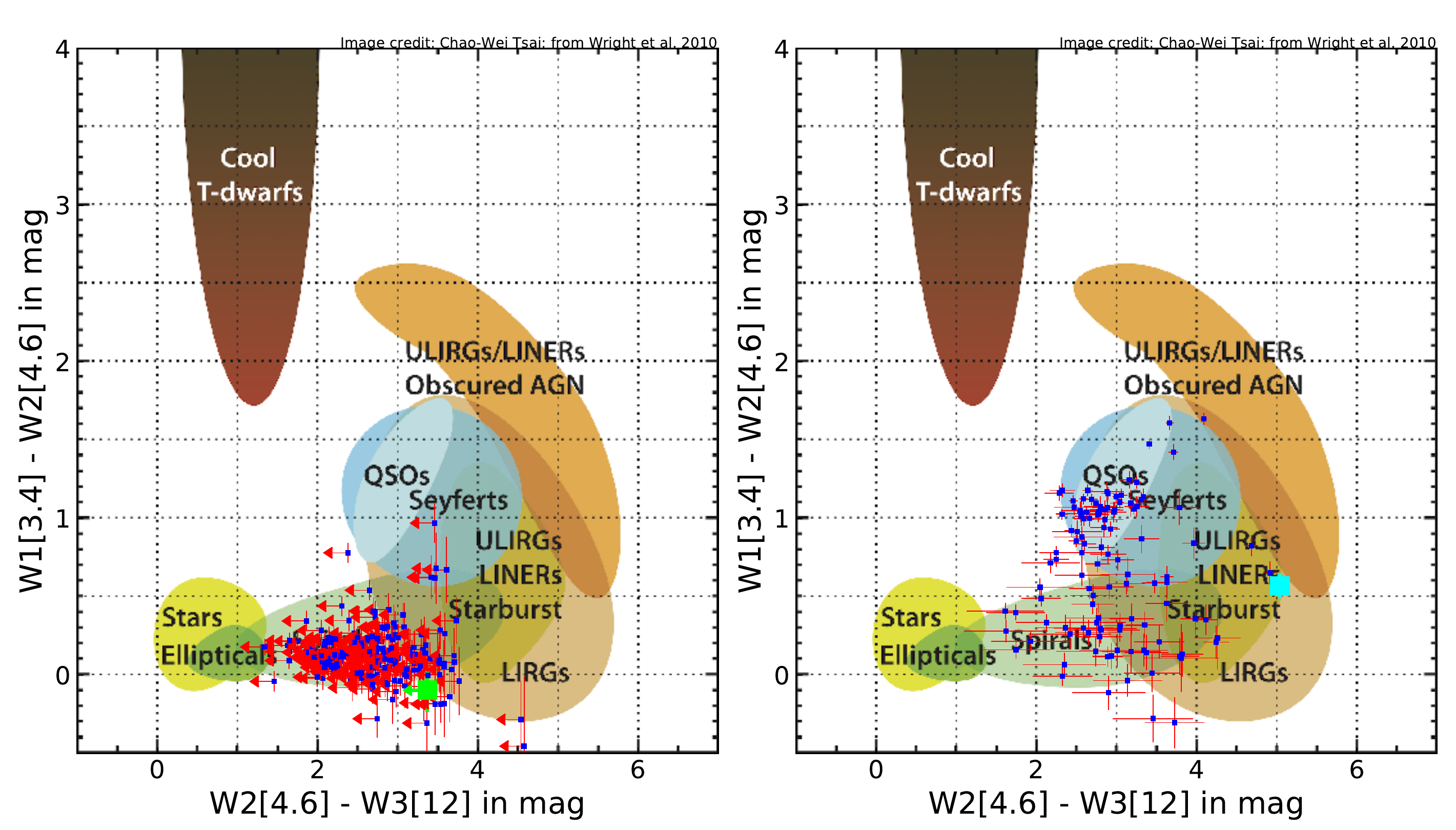}
    \caption{WISE colour-colour diagrams \citep{wright2010} for the 318 objects in Table \ref{table:target-sources-infor} with WISE detections. The left subfigure is for the 203 objects with upper limits on W2-W3, that mainly occupy the `spiral galaxy', `elliptical galaxy', and `star' regions of the plot. The right subfigure shows the 115 objects detected in the W3 band. These are more widely distributed with a relatively higher number density in QSO/Seyfert region. The large green marker in the left subfigure and large cyan marker in the right subfigure mark the sources SDSS J090331+010847 and SDSS J113622+004854 that have been detected in \mbox{H\,{\sc i}} absorption. }
    \label{fig:wise-color-color}
\end{figure*}

\subsection{Sample properties}
We now discuss some of the multi-wavelength properties of the sample of 326 radio sources that we will search for \hi absorption in. Given the sensitivity of our 2 hr and 6 hr observations (around 4.6\,mJy beam$^{-1}$ and 2.8\,mJy beam$^{-1}$ per spectral channel respectively), we can divide the sample as follows: 
\begin{itemize}
    \item 
{\bf The bright sub-sample: } This is a sub-sample of 84 sources with peak flux densities above 45\,mJy beam$^{-1}$ at 855.5\,MHz. For a 45\,mJy source, a strong \hi absorption line with a peak optical depth $\tau=0.26$
 and velocity width $120$\,km s$^{-1}$ should be detectable at the $5\sigma$ level in a 2\,hr ASKAP observation\footnote{A peak optical depth of $\tau=0.26$ roughly corresponds to the highest value seen for associated lines in the large \cite{maccagni2017} sample, while a velocity width of  $120$\,km s$^{-1}$ is the value assumed by \cite{allison2022} for associated absorption lines. }
 and so 40-50\,mJy is considered to be the effective detection limit for a 2\,hr FLASH survey \citep{allison2022}. 
For sources brighter than 45\,mJy, lines with correspondingly lower optical depth would be detectable at the same confidence level. 
\item
{\bf The faint sub-sample:} This comprises the remaining 242 sources with peak flux densities between 10 and 45\,mJy beam$^{-1}$ at 855.5\,MHz. These weaker sources were included mainly to search for \hi absorption in the deeper 6 hr GAMA fields. For a 10\,mJy source, an \hi absorption line with velocity width $120$\,km s$^{-1}$ would need to have an optical depth of $\tau\sim1.0$ to be detectable at $5\sigma$ level in a 6\,hr GAMA observation.  
\end{itemize}

\begin{figure*}
	\includegraphics[width=16cm]{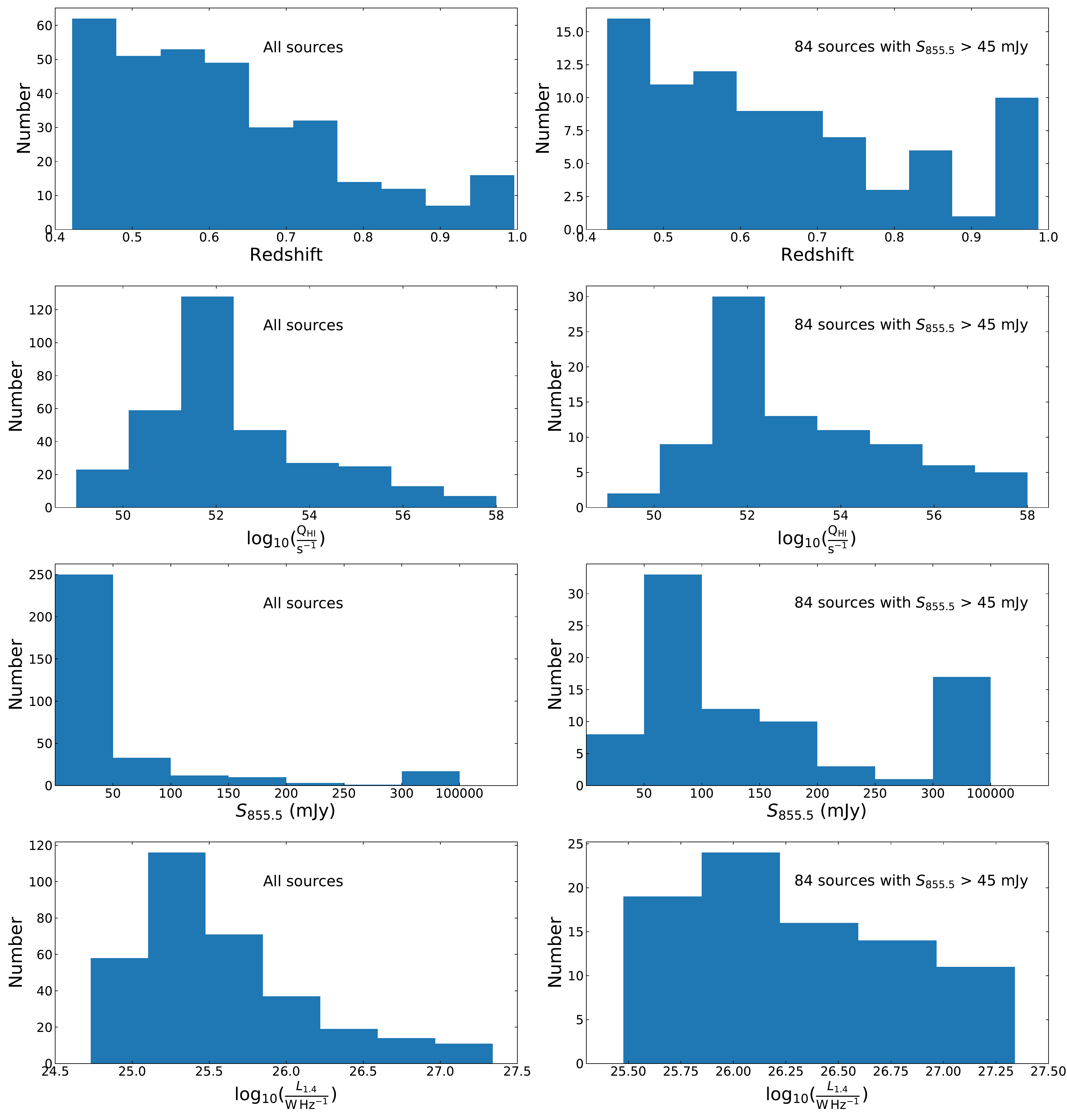}
    \caption{Some general properties of our sample. The left plots, from top to bottom, are redshift, ionising photon rate, flux density $S_{855.5}$, and rest frame radio power at 1.4 GHz, $L_{1.4}$, distributions for our entire sample. The right plots are the same as the left-hand plots but only include sources from the bright subsample with $S_{855.5}$ above 45 mJy.} 
    \label{fig:sample_properties} 
\end{figure*}

\subsubsection{General properties}
Table \ref{table:target-sources-infor} lists some basic information for our targets.
The weakest source used in our \hi absorption search has a peak flux density $S_{855.5}$ of 10 mJy and the brightest source is 3.9\,Jy. 
Fig. \ref{fig:sample_properties} shows some general properties of the sample. While the mid-point of our ASKAP observing band corresponds to $z=0.71$ for \hi, the median redshift of the sample in Table \ref{target_radio_sources_sample} is 0.59. The under-representation of objects with $0.7<z<1.0$ in our sample arises mainly from the optical redshift limits of the SDSS and GAMA redshift surveys. 

The rest-frame radio power at 1.4 GHz, $L_{1.4}$, is calculated from the ASKAP integrated flux density and spans a range from $10^{24.73}$ W\,Hz$^{-1}$ to $10^{27.34}$ W\,Hz$^{-1}$. From  visual inspection of the FIRST radio morphology of the 326 sources, 248 sources are compact ($\leq$ 30 kpc in size at $z\sim 0.5$) and 78 sources are extended (with a typical size of 180 kpc at $z\sim0.5$).  

\subsubsection{Optical spectroscopic properties}
\label{optical_properties}
Based on the spectroscopic classifications in \cite{ching2017}, we find that our targets include 153 low-excitation radio galaxies (LERGs, see \cite{heckman2014}), 32 high-excitation radio galaxies (HERGs) and 36 broad emission-line objects/QSOs (class AeB). There is one BL Lac object (SDSS J142526.18-011825.74) and three objects where the spectrum is classified as `unusual'. The remaining objects lack a spectroscopic classification in the \cite{ching2017} paper. Of the objects with existing spectroscopic classifications, the majority (69\%) are LERGs.

\subsubsection{Mid-IR properties from WISE}
We compiled the mid-infrared properties of our targets from the  Wide-field Infrared Survey Explorer (WISE; \citealt{wright2010}) catalogue. WISE is an all-sky survey operated at mid-infrared bands of 3.4, 4.6, 12, and 22 $\upmu$m. 

This was done by cross-matching the SDSS coordinates of our targets with the ALLWISE catalogue \citep{cutri2013} using a 5-arcsec matching radius.  Of the 326 sources in Table \ref{table:target-sources-infor}, 318 sources have a WISE counterpart in the W1 (3.4$\,\mu$m) band. This high matching rate is what we would expect for radio-loud AGN in our target redshift range, since these objects are usually massive galaxies with an old stellar population that emits strongly at wavelengths near 2-3$\mu$m \cite[e.g.][]{vanbreugel1998}. Of these 318 sources, all were detected in both the W1 (3.4$\,\mu$m) and W2 (4.6$\,\mu$m) bands and 115 were also detected in the W3 (12$\,\mu$m) band. 

The WISE colour-colour diagram \citep{wright2010} is often used to separate various classes of objects, see Fig. \ref{fig:wise-color-color}. We were able to measure both W1-W2 and W2-W3 colours for 115 of the objects in Table \ref{table:target-sources-infor}, and upper limits on W2-W3 for a further 203 objects. The 203 objects with upper limits on W2-W3 could plausibly lie in either the `spiral galaxy' or the  `elliptical galaxy' region in Fig. \ref{fig:wise-color-color}. What is interesting is that almost all of them have W1-W2 $<0.8$ mag which means the mid-IR light comes mainly from stars \citep{stern2012}. The 115 objects detected in the W3 band are more widely distributed with a relatively higher number density at QSO/Seyfert region.  

Of the 318 objects, 53 (17\%) have W1-W2 > 0.8, including 5 HERGs, placing them in the AGN region of the WISE diagram. 52 of these 53 objects have been detected in W3. This high fraction is not surprising, as W1-W2 $>0.8$ has been proved to be a effective criterion in selecting AGN candidates \citep{stern2012}. Sources with W1-W2 > 0.8 have a high probability of hosting an AGN with a high accretion rate, and the reddening of optical light from central AGN makes them more easily detected in WISE W3 band. 

\subsubsection{UV properties from GALEX} 
We investigated the UV properties of our targets by cross-matching their SDSS coordinates with Galaxy Evolution Explorer (GALEX) catalogue \citep{bianchi2017} using a 3 arcsec matching radius. We then made a power-law fit to the photometric data of SDSS bands and GALEX bands, if available, after applying Galactic extinction \citep{schlegel1998} and shifting to the rest frame. Finally we calculated the ionising photon rate by following the method described in \cite{curran2017}, which is listed in Table  \ref{table:target-sources-infor}. 

Most of our targets have an ionising photon rate below $1.7\times10^{56}$\,s$^{-1}$, the critical value above which it is argued that all the neutral hydrogen in host galaxies could be ionised (\citealt{curran2008,curran2012,curran2017}), see Fig.  \ref{fig:sample_properties}. 

\begin{landscape}
\begin{table} 
\centering
\caption{Our target radio sources with known redshift between 0.42 to 1.00 and $S_{855.5}\geq10 $ mJy in the GAMA fields. Column [1] is island IDs. Column [2] and [3] are RA and Dec. of islands. Column [4] is the component designation within each island. Column [5] and [6] are RA and Dec. of components. Column [7] is peak flux density, $S_{855.5}$. Column [8] is the cross-matched SDSS ID.  Column [9] is the optical redshift of the host galaxy of each radio source. Column [10] indicates whether the optical source is in the \protect\cite{ching2017} sample. Column [11] lists the separation between island coordinates and optical coordinates. We show here a part of the Table to illustrate its content and format. For the whole table, please refer to the online material.}
\label{table:target-sources}
\begin{tabular}{lcccccccccc}
\hline
Island ID &  \multicolumn{2}{c}{Island coordinates}  & Comp\_ID  &   \multicolumn{2}{c}{Component coordinates} & Peak flux density &SDSS ID  & Redshift & In \cite{ching2017}& Sep. \\
& RA\,[J2000] & Dec.\,[J2000] &    &RA\,[J2000]  & Dec.\,[J2000] &  mJy\,beam$^{-1}$ & & & & arcsec \\

[1]& [2] & [3] & [4] & [5] & [6] & [7] & [8] & [9]& [10]& [11]\\

\hline
\multicolumn{11}{c}{Islands that contain one component}\\
SB13285\_island\_905	&08:35:14.06	&+00:16:13.04	&a	&08:35:14.05	&+00:16:12.98	&13.5	&J083514.23+001614.71	&0.574$^{s}$	&yes	&3.09\\
SB13285\_island\_788	&08:36:26.34	&+03:40:54.81	&a	&08:36:26.34	&+03:40:54.71	&15.7	&J083626.33+034055.31	&0.663$^{s}$	&no	&0.53\\
SB13285\_island\_1014	&08:37:10.76	&+02:50:46.82	&a	&08:37:10.74	&+02:50:46.63	&11.5	&J083710.75+025046.88	&0.543$^{g}$	&yes	&0.19\\
SB13285\_island\_675	&08:37:22.77	&+02:21:01.89	&a	&08:37:23.18	&+02:21:03.12	&18.0	&J083722.96+022104.65	&0.606$^{g}$	&yes	&3.94\\
SB13285\_island\_531	&08:37:42.21	&+00:00:47.07	&a	&08:37:42.18	&+00:00:46.36	&24.7	&J083742.25+000045.56	&0.455$^{g}$	&yes	&1.62\\
SB13285\_island\_138	&08:38:11.76	&-01:59:35.49	&a	&08:38:11.76	&-01:59:35.72	&88.9	&J083811.77-015935.40	&0.561$^{g}$	&no	&0.20\\
SB13285\_island\_201	&08:38:34.59	&+00:02:08.21	&a	&08:38:34.58	&+00:02:07.89	&65.0	&J083834.58+000208.24	&0.608$^{g}$	&yes	&0.11\\
SB13285\_island\_575	&08:38:53.25	&+02:38:06.58	&a	&08:38:53.42	&+02:38:04.05	&22.6	&J083853.48+023802.92	&0.462$^{g}$	&yes	&5.04\\

\multicolumn{11}{c}{......}\\

\multicolumn{11}{c}{Islands that contain multi components}\\
SB13285\_island\_2	&08:36:20.41	&-00:37:23.20	&a	&08:36:20.38	&-00:37:25.36	&467.4	&J083620.43-003723.65	&0.526$^{g}$	&yes	&0.60\\
SB13285\_island\_2	&08:36:20.41	&-00:37:23.20	&b	&08:36:20.45	&-00:37:22.16	&431.7	&J083620.43-003723.65	&0.526$^{g}$	&yes	&0.60\\
SB13285\_island\_2	&08:36:20.41	&-00:37:23.20	&c	&08:36:20.44	&-00:36:58.48	&13.1	&J083620.43-003723.65	&0.526$^{g}$	&yes	&0.60\\
SB13285\_island\_47	&08:38:14.27	&-00:26:00.77	&a	&08:38:14.39	&-00:26:01.72	&115.4	&J083814.28-002600.92	&0.460$^{g}$	&yes	&0.19\\
SB13285\_island\_47	&08:38:14.27	&-00:26:00.77	&b	&08:38:14.11	&-00:26:00.43	&93.7	&J083814.28-002600.92	&0.460$^{g}$	&yes	&0.19\\
SB13285\_island\_150	&08:45:13.11	&-00:43:13.71	&a	&08:45:13.34	&-00:42:57.57	&81.1	&J084512.97-004318.22	&0.701$^{g}$	&yes	&4.99\\
SB13285\_island\_150	&08:45:13.11	&-00:43:13.71	&b	&08:45:12.72	&-00:43:41.02	&48.4	&J084512.97-004318.22	&0.701$^{g}$	&yes	&4.99\\
SB13285\_island\_305	&08:47:00.63	&+00:37:16.78	&a	&08:47:00.40	&+00:37:15.62	&40.3	&J084700.78+003718.36	&0.428$^{s}$	&no	&2.77\\
\multicolumn{11}{c}{......}\\
\hline
\end{tabular}
\\
\medskip
$^{s}$redshift from SDSS.
$^{g}$redshift from GAMA.
$^{n}$redshift from NED. 
$^{c}$ redshift from \cite{ching2017}.
\end{table}
\end{landscape}

\begin{landscape}
\begin{table}

\centering

\caption{Basic information for our target sources. Column [1] is the SDSS optical ID. Column [2] is redshift. Column [3] is luminosity distance. Column [4] and Column [5] are the NVSS and ASKAP integrated flux densities. Column [6] is rest frame radio power at 1.4 GHz calculated from Column [5]. Column [7] is the  spectral index at 855.5-1400\,MHz calculated from the NVSS and ASKAP flux densities. Column [8], [10], and [12] list the W1, W2, and W3 magnitudes from WISE, with their errors listed in Columns [9], [11], and [13]. Column [14] is the ionising photon rate. Column [15] denotes the activity type reported by \protect\cite{ching2017}. Column [16] is the radio morphology classification based on our visual inspection of FIRST images, where C means `Compact' and Ext means `Extended'. Column [17] indicates GAMA fields. Column [18] indicates the SBID from which we extracted each spectrum. Column [19] is the rms spectral noise in a single 18.5 kHz channel. Column [20] is the peak optical depth $\tau_{\rm pk}$. For the non-detections, a 3$\sigma$ limit of $\tau_{\rm pk}$ based on spectral noise from Column [19] is given. Column [21] is the estimated \mbox{H\,{\sc i}} column density. For non-detections in \hi, we assume a Gaussan absorption line with a FWHM of 120 km\,s$^{-1}$ and the upper limit for $\tau_{\rm pk}$ from Column[20]. We show here a part of the table to regard its content and format. For the whole table, please refer to the online version. }
\label{table:target-sources-infor}
\scalebox{0.7}{
\begin{tabular}{lcccccccccccccccccccc}
\hline 
SDSS ID & Redshift & D & $S_{\rm NVSS}$ & $S_{\rm ASKAP}$  & $\log_{10} \left(\frac{L_{1.4\,\rm GHz}}{{\rm W}\,{\rm Hz}^{-1}}\right)$ & $\alpha$ & $W1$ & $\pm$  & $W2$ & $\pm$ & $W3$ & $\pm$  & $\log_{10} \left(\frac{Q_{HI}}{s^{-1}}\right)$ & Type & Structure &Field &SBID & rms & $\tau_{\rm pk}$ & $\log_{10} \left(\frac{N_{\rm HI}}{\rm cm^{-2}}\right)$ \\
& $z$ & Mpc & mJy & mJy  &  & $S_{\nu}\sim \nu^{\alpha}$ & mag & mag & mag & mag & mag & mag& & & & & & mJy\,chan$^{-1}$ & &\\

[1]& [2] & [3] & [4] & [5] & [6] & [7] & [8] & [9] & [10]& [11]& [12]& [13]& [14]& [15]& [16]& [17]& [18]& [19]& [20]& [21]\\
\hline
J083514.23+001614.71	&0.574	&3345.64	&9.7	&15.53	&25.12	&-0.96	&15.000	&0.035	&14.932	&0.072	&>12.797	&...	&52.15	&...	&C	&G9A	&13293	&3.28	&<1.247	&<22.46\\
J083620.43-003723.65	&0.526	&3010.90	&560.0	&876.93	&26.79	&-0.91	&16.821	&0.105	&16.139	&0.202	&>12.660	&...	&53.49	&HERG	&C	&G9A	&13293	&2.89	&<0.012	&<20.45\\
J083626.33+034055.31	&0.663	&3986.27	&14.3	&17.83	&25.31	&-0.45	&14.741	&0.032	&14.751	&0.066	&>11.888	&...	&52.56	&...	&C	&G9A	&13293	&4.89	&<2.414	&<22.75\\
J083710.75+025046.88	&0.543	&3128.55	&9.6	&13.99	&25.03	&-0.76	&15.038	&0.035	&14.689	&0.061	&>12.296	&...	&53.03	&LERG	&C	&G9A	&13293	&3.68	&<2.520	&<22.77\\
J083722.96+022104.65	&0.606	&3573.09	&21.0	&37.25	&25.55	&-1.16	&13.140	&0.043	&13.180	&0.049	&>11.722	&...	&50.14	&LERG	&Ext	&G9A	&13293	&3.36	&<0.963	&<22.35\\
J083742.25+000045.56	&0.455	&2530.80	&29.3	&25.42	&25.13	&0.29	&15.870	&0.052	&15.276	&0.088	&11.646	&0.239	&51.78	&LERG	&C	&G9A	&13293	&2.54	&<0.366	&<21.93\\
J083811.77-015935.40	&0.561	&3254.21	&72.5	&89.65	&25.86	&-0.43	&14.755	&0.032	&14.684	&0.059	&>12.185	&...	&53.4	&...	&C	&G9A	&13293	&3.29	&<0.117	&<21.43\\
J083814.28-002600.92	&0.46	&2563.99	&138.0	&209.42	&26.05	&-0.85	&16.053	&0.060	&15.932	&0.159	&>12.435	&...	&49.45	&LERG	&C	&G9A	&13293	&2.62	&<0.040	&<20.97\\
J083834.58+000208.24	&0.608	&3587.42	&42.0	&67.28	&25.81	&-0.96	&15.023	&0.035	&14.947	&0.067	&>12.491	&...	&50.27	&LERG	&C	&G9A	&13293	&2.88	&<0.141	&<21.52\\
J083853.48+023802.92	&0.462	&2577.30	&20.3	&38.44	&25.32	&-1.3	&15.512	&0.043	&15.353	&0.098	&12.346	&0.460	&49.44	&LERG	&Ext	&G9A	&13293	&3.10	&<0.582	&<22.13\\
J083929.72-005829.96	&0.555	&3212.20	&10.9	&17.76	&25.15	&-0.99	&14.715	&0.033	&14.531	&0.057	&>12.601	&...	&50.25	&LERG	&C	&G9A	&13293	&2.91	&<0.816	&<22.28\\
J084002.46+003013.59	&0.427	&2346.71	&22.7	&37.19	&25.23	&-1.0	&15.007	&0.035	&14.838	&0.070	&>12.027	&...	&52.21	&LERG	&C	&G9A	&13293	&2.54	&<0.246	&<21.76\\
J084026.37+000006.75	&0.980	&6444.69	&16.0	&19.68	&25.69	&-0.42	&15.004	&0.035	&13.932	&0.039	&11.462	&0.247	&56.01	&AeB	&C	&G9A	&13293	&3.88	&<1.009	&<22.37\\
J084225.51+025252.72	&0.425	&2333.68	&16.7	&16.89	&24.89	&-0.02	&14.488	&0.030	&14.152	&0.046	&12.027	&0.380	&54.27	&LERG	&C	&G9A	&13293	&2.57	&<0.666	&<22.19\\
J084253.77+031427.91	&0.519	&2962.75	&12.8	&18.99	&25.12	&-0.8	&15.096	&0.038	&15.155	&0.087	&>12.521	&...	&52.02	&...	&C	&G9A	&13293	&2.81	&<0.828	&<22.28\\
J084428.51+014818.77	&0.868	&5548.03	&15.1	&29.05	&25.76	&-1.33	&15.296	&0.038	&14.560	&0.055	&11.546	&0.199	&55.3	&AeB	&C	&G9A	&13293	&3.20	&<0.522	&<22.08\\
J084453.56-004208.47	&0.442	&2444.95	&11.1	&16.98	&24.93	&-0.86	&15.089	&0.036	&14.685	&0.057	&>11.968	&...	&51.31	&LERG	&C	&G9A	&13293	&2.33	&<0.954	&<22.35\\
J084512.97-004318.22	&0.701	&4267.15	&109.8	&163.77	&26.32	&-0.81	&15.290	&0.039	&14.509	&0.052	&>12.131	&...	&54.97	&AeB	&Ext	&G9A	&13293	&2.69	&<0.795	&<22.27\\
J084543.95+005736.97	&0.427	&2346.71	&157.8	&203.13	&25.97	&-0.51	&15.145	&0.038	&14.829	&0.072	&11.785	&0.292	&52.43	&LERG	&C	&G9A	&13293	&2.18	&<0.034	&<20.90\\
J084553.45-024100.15	&0.468	&2617.30	&54.9	&67.85	&25.58	&-0.43	&12.614	&0.024	&11.516	&0.021	&8.773	&0.028	&57.14	&...	&C	&G9A	&13293	&6.81	&<0.353	&<21.91\\
J084615.46+021141.12	&0.651	&3898.46	&39.6	&61.66	&25.83	&-0.9	&14.971	&0.035	&14.899	&0.070	&>12.661	&...	&52.24	&LERG	&C	&G9A	&13293	&2.57	&<0.197	&<21.66\\
J084620.68+020228.07	&0.498	&2819.35	&10.2	&19.0	&25.08	&-1.26	&15.271	&0.038	&14.886	&0.068	&>11.820	&...	&50.64	&HERG	&C	&G9A	&13293	&2.26	&<0.458	&<22.03\\
J084623.12-012218.41	&0.446	&2471.30	&7.8	&11.0	&24.75	&-0.7	&15.147	&0.035	&15.086	&0.076	&>12.120	&...	&51.52	&...	&C	&G9A	&13293	&2.39	&<1.105	&<22.41\\
J084700.78+003718.36	&0.428	&2353.24	&41.4	&60.72	&25.45	&-0.78	&15.305	&0.038	&14.666	&0.059	&12.097	&0.336	&50.46	&...	&Ext	&G9A	&13293	&2.32	&<0.183	&<21.63\\
J084709.97+005459.33	&0.430	&2366.29	&45.8	&72.73	&25.53	&-0.94	&15.264	&0.040	&15.104	&0.094	&>12.592	&...	&52.09	&LERG	&C	&G9A	&13293	&2.17	&<0.104	&<21.38\\
\hline 
\end{tabular}}
\\
\medskip
\end{table}
\end{landscape}

\begin{table*}
\caption{The fitted parameters for the two associated \mbox{H\,{\sc i}} absorption lines detected in our search: Column [1] lists the SDSS name and Column [2] the optical redshift. Column [3] is the number of fitted Gaussian profiles. Column [4] is offset between the \hi velocity and the optical systemic velocity. Column [5] is the FWHM for the line as measured by the FLASHfinder and Column [6] is the peak optical depth. Column [7] is the integrated optical depth. Column [8] is the estimated \mbox{H\,{\sc i}} column density assuming a spin temperature of 100\,K and covering factor of 1. Column [9] lists the SBID reference number of the observation from which the corresponding measurements come.}
\centering
\begin{tabular}{lcccccccc}
\hline
Name & Redshift &$N_{\rm Gau}$ & $V_{\rm peak}$ &$V_{\rm FWHM}$ & $\tau_{\rm pk}$ & $\tau_{\rm int}$ & $\log_{10} \left(\frac{N_{\rm HI}}{\rm cm^{-2}}\right)$ & SBID \\
& $z$ & & $\rm km\,s^{-1}$ &$\rm km\,s^{-1}$ & & $\rm km\,s^{-1}$ &  & \\
\relax
[1]&[2]&[3]&[4]&[5]&[6]&[7]&[8]&[9]\\
\hline
SDSS J090331+010847 &0.522 & 1 & $-33.97^{+1.26}_{-1.31}$ & $66.90^{+3.39}_{-3.19}$ &$1.77^{+0.16}_{-0.16}$& $118.41^{+7.25}_{-7.51}$&$22.33^{+0.03}_{-0.02}$ & 13283\\[5pt]
SDSS J113622+004852 &0.563 & 1 & $47.30^{+1.69}_{-1.64}$ & $52.95^{+5.08}_{-4.47}$ &$0.14^{+0.01}_{-0.01}$& $7.40^{+0.54}_{-0.53}$&$21.13^{+0.03}_{-0.03}$ & 13306\\[5pt]
 && 1 & $45.65^{+3.18}_{-3.09}$ & $51.37^{+7.59}_{-6.36}$ &$0.16^{+0.02}_{-0.02}$ & $8.25^{+1.11}_{-1.02}$&$21.18^{+0.05}_{-0.06}$ & 13334\\[5pt]
\hline
\end{tabular}
\label{tab:fitted_paras}
\end{table*}

\section{Results}
\subsection{Line-finder results}
We used the Bayesian method developed by \cite{allison2012_2} to detect and parametrize \mbox{H\,{\sc i}} absorption lines. This method uses multi-modal nested sampling, a Monte Carlo sampling algorithm, to simultaneously find and fit any potential \mbox{H\,{\sc i}} absorption line in a radio spectrum. The Bayes factor (e.g. \citealt{kass1995}) $B$ is used to quantify the relative significance of a Gaussian spectral line model versus a noise model. $\ln(\rm B)$ > 1 means a Gaussian spectral line model is preferred over a model with no line, see \cite{allison2012_2} for the details of this method. In a simulation in which 1000 fake absorption lines were randomly produced, this method can find all lines with FWHM $\ge 30 \rm km\,s^{-1}$ and peak signal noise ratio $\ge 5$ in a single 18.5-kHz channel, see Fig. 4 in \cite{allison2020}. This method has also been proved to be effective in finding \mbox{H\,{\sc i}} absorption lines in observed ASKAP data (e.g.  \citealt{allison2020, sadler2020})

We first extracted spectra of our 326 targets, spanning from $-500$\, km\,s$^{-1}$ to $+500$\,km\,s$^{-1}$ in the rest frame and using the post-processed spectra from SBIDs 13293, 13283, 13306, 13335, 13294, and 13273. We then ran the FLASHfinder\footnote{\url{https://github.com/drjamesallison/flash_finder}} on these cut-down spectra. To estimate the number of false-positive absorption features  produced by noise or spectral glitches we then inverted our spectra and ran the FLASHfinder on them again. Lines due to noise rather than a genuine signal should produce similar distributions of features in both absorption and emission. For details of this method, please refer to \cite{allison2021}. In this process, we detected 23 negative features with $\ln(B)>1$ in our normal spectra and 18 negative features with $\ln(B)>1$ in the inverted spectra.

\begin{figure*}
	\includegraphics[width=17.5cm]{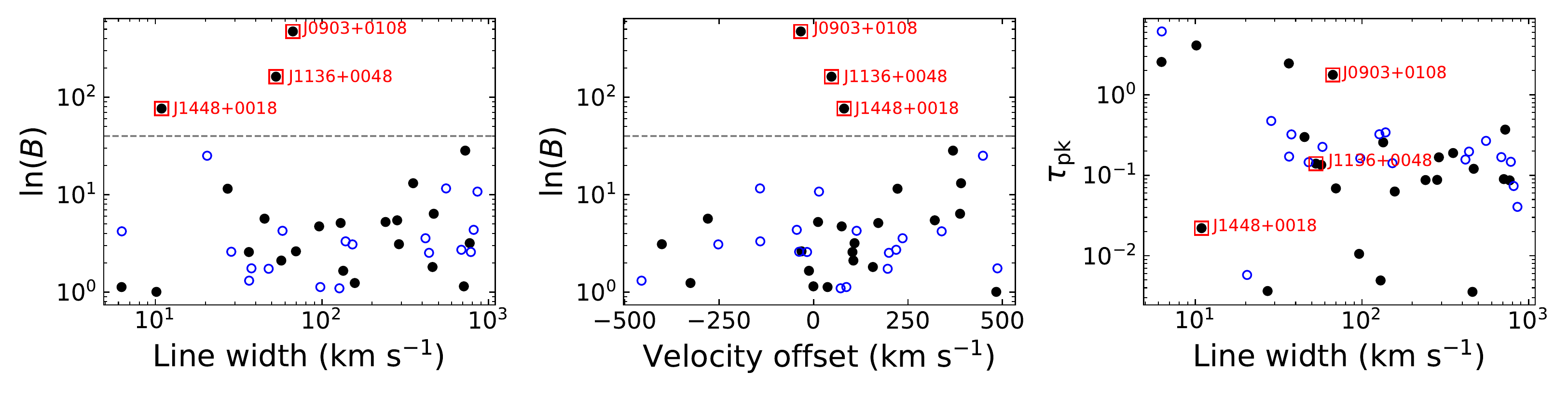}
    \caption{The FLASHfinder results for our 326 targets from SBID observations of 13293, 13283, 13306, 13335, 13294, and 13273. We detect 23 negative features, indicated by black filled circles, with $\ln(B)$ > 1 in our normal spectra and 18 negative features, indicated by blue open circles, with $\ln(B)$ > 1 in our inverted spectra. The $\ln(B)$ denotes the significance of a negative feature; the line width is FWHM; the velocity offset is the velocity where the fitted $\tau_{\rm pk}$ is with respect to the systemic velocity and the $\tau_{\rm pk}$ is the peak optical depth. Three sources with high value of $\ln(B)$ have been marked with red squares. The horizontal dashed lines in the left and middle plots are at $\ln(B)$ = 40.} 
    \label{fig:line_finder_result} 
\end{figure*}

To distinguish the real detections from false features, we plotted the FLASHfinder results in various parameter spaces as shown in Fig. \ref{fig:line_finder_result}. 
Three sources, labelled as J0903+0108, J1136+0048, and J1448+0018 in Fig. \ref{fig:line_finder_result}, with value of $\ln(B)$ $>$ 40 stand out in the $\ln(B)$ vs line width and $\ln(B)$ vs velocity offset plots. However, these three sources are clustered with others in the $\tau_{\rm pk}$ vs line width plot that was previously used to identify real detections in ATCA observations of compact radio sources \citep{allison2012}. 

As a final step, we visually inspected the 326 radio spectra including those not reported with $\ln(\rm B)$ > 1 and confirmed the accurate reports from the FLASHfinder. We further confirmed that J0903+0108 and J1136+0048 are genuine detections, whereas the narrow J1448+0018 line is a spectral glitch at the edge of an ASKAP beam-forming interval (we plotted the edges of beam-forming intervals on the FLASH spectra). 

\begin{figure*}
	\includegraphics[width=17.5cm]{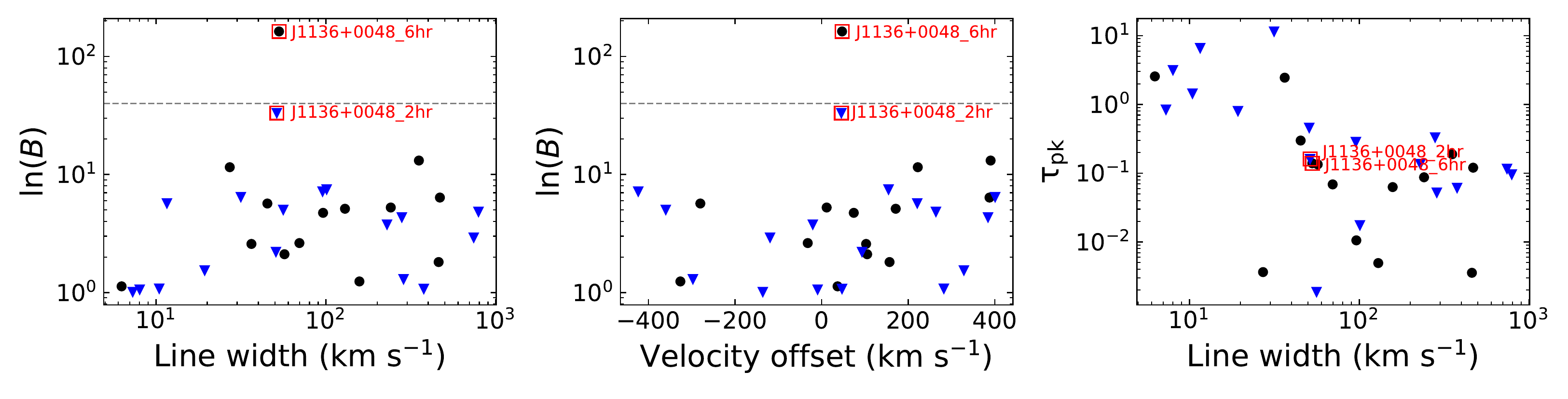}
    \caption{The comparison of FLASHfinder results between 2 hours observation and 6 hours observations. Although we observed each GAMA field twice, only GAMA\_09A, GAMA\_12A and GAMA\_15A have post-processed spectra from both the 2 hours and 6 hours observations, see Table \ref{table:obs_summary}. The black filled circles are for 6 hours observations with SBID 13293, 13306, and 13294 and the blue filled triangles are for 2 hours observations with SBID 13285, 13334 and 13336. The horizontal dashed lines in the left and middle plots are at  $\ln(B)$ = 40.} 
    \label{fig:finder_compa} 
\end{figure*}

\subsection{Comparison of the 6hr and 2hr data} 
We have both 6 hour and 2 hour post-processed spectra for the  GAMA\_09A, GAMA\_12A and GAMA\_15A fields. As a comparison, we show the results of FLASHfinder of these fields in Fig. \ref{fig:finder_compa}. In total, we detected 14 and 17 negative features with $\ln(B)$ > 1 in the 6 hour and 2 hour spectra of the three fields respectively. The absorption line in SDSS J1136+0048 has $\ln(B)$ = 162.8 and $\ln(B)$ = 33.2 in the 6 hour and 2 hours observations respectively. Although the $\ln(B)$ value in SDSS J1136+0048 for the 2 hour observation is lower than in the 6 hour observation, it is still a significant detection. Fig. \ref{fig:finder_compa} shows that, apart from  SDSS J1136+0048, there is no other negative feature that stands out in $\ln(B)$. The number of spurious detections in the 6 hour observations is 13, compared to 16 in the 2 hour observation.   The distributions of spurious negative features are similar in the 2 hour and 6 hour observations, consistent with these being random noise peaks.

\subsection{Detection of associated \mbox{H\,{\sc i}} absorption in SDSS J090331+010847.}

\subsubsection{The \hi spectrum}
We detected an associated \mbox{H\,{\sc i}} absorption line towards the radio galaxy SDSS J090331+010847 at redshift $z=0.522$ \citep{ching2017}. We show the full-band radio spectrum for this object (from SBID 13283) in Fig. \ref{fig:J090331+010847_full_band_spectrum}. The quality of the spectrum is good and the band is free of radio frequency interference (RFI), though a few glitches are seen, notably between 916 and 920\,MHz. These are caused by a `correlator dropout', an intermittent ASKAP issue in which some of the  high-resolution frequency channels in a single correlator block cease to provide data.

An absorption line is seen at 933.352 MHz which is consistent with the GAMA optical redshift of SDSS J090331+010847. The measured average rms noise per channel around 933.352 MHz is 4.59 mJy\,beam$^{-1}$.  An enlarged spectrum of the line is shown in Fig. \ref{fig:optical_radio_plot} and the line is best fitted with a single Gaussian component via a Bayesian approach as described by \cite{allison2015}. The fitted parameters for velocity width, peak optical depth, integral optical depth, and FWHM are listed in Table \ref{tab:fitted_paras}. The \mbox{H\,{\sc i}} column density is given by: 
\begin{equation}\label{equation:column_density}
N_{\rm HI} = 1.82\times10^{18}\,T_{\rm s}\times\int \tau(V) {\rm d}{V} 
\end{equation}

For J090331+010847, this yields a value of N$_{\rm HI}$ =  
$2.14^{+0.15}_{-0.10}\times10^{22} $ atoms\,cm$^{-2}$ which is much higher than the lowest \mbox{H\,{\sc i}} column density used to define Damped Lyman Alpha absorbers (DLA; A DLA is defined to have an \mbox{H\,{\sc i}} column density above $2\times10^{20}$ atoms cm$^{-2}$).
 In Eq. \ref{equation:column_density}, $T_{s}$ is the spin temperature of the \mbox{H\,{\sc i}} gas, V is the velocity in the rest frame and $\tau(V)$ is the optical depth given by 
 \begin{equation}\label{equation:tau}
 \tau(V) = -\ln[1+\Delta S/(C_{\rm f} S_{\rm cont})].
 \end{equation}
 Here, $C_{\rm f}$ is the covering factor indicating the fraction of background radio continuum intercepted by the foreground \mbox{H\,{\sc i}} gas, $\Delta S$ is the flux density in the line\footnote{For an absorption line, the value of $\Delta S$ is negative in equation \ref{equation:tau} } and $S_{\rm cont}$ is the continuum flux density. 
 From the expression for $\tau(V)$ we can see that it is indeed independent of redshift. 

In our calculation, we assumed a covering factor of 1 and  adopted a spin temperature of 100 K. The covering factor and spin temperature are two uncertain parameters, which can influence the calculated \mbox{H\,{\sc i}} column density. 
In some very long baseline interferometry (VLBI) \mbox{H\,{\sc i}} spectral observations where the absorbing \mbox{H\,{\sc i}} has been resolved against its background radio continuum, the covering factor is generally less (or much less) than 1. For example, in the case of Cygnus\,A, the \mbox{H\,{\sc i}} gas was detected only against about half of the central radio continuum \citep{struve2010}. Similar situations have been also reported in other VLBI studies (e.g. \citealt{morganti2013,schulz2021}). 

The \mbox{H\,{\sc i}} spin temperature we adopted is also a fiducial value often used when a measurement of it is unavailable. In reality, the spin temperature may be higher than 100 K. Previously known measurements of \mbox{H\,{\sc i}} spin temperature towards extra-galactic targets and those in our own Milky Way galaxy show that 100\,K is at the low end of the distribution of \mbox{H\,{\sc i}} spin temperatures (e.g. \citealt{reeves2015,reeves2016,kanekar2014,murray2018,allison2021}). Additionally, the \mbox{H\,{\sc i}} spin temperature may change dramatically across regions in a galaxy. In the circumnuclear region of AGN, it could be as high as $\sim$ $10^{4}$ K \citep{bahcall1969}. Therefore, the \mbox{H\,{\sc i}} column density we calculated based on Eq. \ref{equation:column_density} for SDSS J090331+010847 should be considered as only a lower limit.

\subsubsection{Optical properties} 
Based on its optical colours, SDSS J090331+010847 is a red galaxy. Fig. \ref{fig:optical_radio_plot} shows ASKAP radio contours overplotted on a $grz$ composite image from the DESI Legacy Imaging Surveys \citep{dey2019} . The GAMA optical spectrum of SDSS J090331+010847 gives a redshift of 0.522 with quality $Q$ = 4, corresponding to a systemic velocity of 156,600 km\,s$^{-1}$, see Fig. \ref{fig:GAMA_spectra}. There are no prominent emission lines in the spectrum but some stellar absorption lines are visible and \cite{ching2017} classified this object as a LERG. 

We fitted the UV-optical-infrared spectral energy distribution (SED) of SDSS J090331+010847 using the \prospect code \citep{robotham2020} to simultaneously fit a stellar model, attenuation model, dust emission model, and Fig. \ref{fig:optical_sed_fit} shows the fitted SED. From this fit, we estimate a stellar mass of $1.82^{+0.37}_{-0.31}\times10^{11} \rm M_{\odot}$, a metallicity $Z = 0.0358^{+0.0143}_{-0.0138}$, a dust mass of $2.08^{+6.10}_{-1.43}\times$ $10^{8} \rm M_{\odot}$, and a star formation rate of $1.47^{+2.56}_{-1.47} \rm M_{\odot}\,yr^{-1}$.  

The SED fit implies that SDSS J090331+010847 had a very high star formation rate early in its history, peaking above 100 $M_{\odot}\,yr^{-1}$. However, the star formation rate decreased dramatically between $z\sim1$ and $z=0.52$ and star formation now appears to be largely quenched in this galaxy. 

\subsubsection{Radio properties}
We explore the radio properties of SDSS J090331+010847 by combining the data from our own observation, FIRST, NVSS, VLA Sky Survey \citep[VLASS,][]{gordon2020}, and other literature measurements. 

With ASKAP we observed SDSS J090331+010847 twice (SBID 13283 and 11068, see Table \ref{table:obs_summary}) and measured an unresolved radio source with flux densities of 56.9 and 55.1 mJy respectively. SDSS J090331+010847 is also compact in FIRST, NVSS and VLASS, with flux densities of 72.69, 63.1, and 38.3 mJy respectively. We note that the NVSS integrated flux density is 9.59 mJy lower than the FIRST flux density, which is unexpected as NVSS has a larger beam and is more sensitive to diffuse radio components. The most likely interpretation is that SDSS J090331+010847 hosts a variable radio source, which is not uncommon for AGN. 

We added some additional radio continuum data points from the literature and fitted the radio spectrum for SDSS J090331+010847 using the equation:

\begin{equation}\label{equation:radio_sed_fit_equ}
S_{\nu}=\frac{S_{\rm max}}{1-e^{-1}}\times\left(\frac{\nu}{\nu_{\rm max}}\right)^{\alpha_{\rm thick}}\times [(1-e^{-(\frac{\nu}{\nu_{\rm max}})^{\alpha_{\rm thin}-\alpha_{\rm thick}}}]
\end{equation}
where $S_{\rm max}$, $\nu_{\rm max}$, $\alpha_{\rm thick}$, and $\alpha_{\rm thin}$ are peak flux density, peak frequency, optically thick index, and optically thin index, respectively \citep{snellen1998}. The fitted radio SED is shown in Fig. \ref{fig:radio_sed_fit} and the fitted parameters are listed in Table \ref{tab:radio_sed_fitted_pa}.

The radio SED of SDSS J090331+010847 peaks at 768\,MHz in the observed frame, corresponding to 1167\,MHz in the source rest frame, with an optically thick index of $+2.61$ and optically thin index of $-0.41$. SDSS J090331+010847 is therefore a GPS radio source, indicating that this is a young radio source in a period when the radio jet is in growing but still confined within its host galaxy. 

The fitted spectral index of $+2.61$ below the peak is similar to that of the extreme GPS source PKS\,B0008-421 \citep{callingham2015}, which has the steepest known slope below the turnover. The current low-frequency fit is constrained mainly by a single data point at 325\,MHz, but the source is undetected in a recent LOFAR image (J.R. Callingham and T.W. Shimwell, private communication) with a flux density $<0.8$\,mJy at 144\,MHz. This LOFAR upper limit is consistent with a steep spectrum below the peak (see Fig. \ref{fig:radio_sed_fit}). 
Deeper observations of this source at frequencies below 300\,MHz would be extremely useful. 

\subsubsection{Infrared properties}
The WISE W1 and W2 magnitudes for SDSS J090331+010847 are 15.671$\pm$0.047 and 15.767$\pm$0.132 respectively, with an upper limit of 12.397 in W3.  The location in the WISE colour-colour diagram is shown by a green marker in Fig. \ref{fig:wise-color-color}. Taking into consideration the measurement errors, SDSS J090331+010847 could be located in any any of the Luminous Infrared Galaxy (LIRG), spiral galaxy, and elliptical galaxy regions. 

The W1-W2 colour can reflect the fraction of emission from an AGN \citep{stern2012}, so the low W1-W2 value of SDSS J090331+010847 means that the AGN in SDSS J090331+010847 contributes a small fraction to total emission which is consistent with that in LERGs with radiatively-inefficient accretion, where the dominant emission in the mid-IR comes from the host galaxy.  

\subsubsection{Summary}
Given the high \mbox{H\,{\sc i}} column density estimated from \mbox{H\,{\sc i}} absorption and the stellar mass derived from UV-optical-infrared SED, we conclude that SDSS J090331+010847 is a massive and gas rich galaxy. If we assume an \hix:dust mass ratio of $\sim100$ like that in the Milky Way \citep[e.g.][]{knapp1974}, then the fitted dust mass of $2.39\times10^8$\,M$_\odot$ implies an \hi mass around  $2\times10^{10}$M$_\odot$ for SDSS J090331+010847. If the \hi mass:dust mass ratio is higher at $z\sim0.5$ than in the Milky Way, as suggested by the work of \cite{menard2009}, then the estimated \hi mass will be somewhat higher. SDSS J090331+010847 therefore appears to be a massive, red, gas-rich galaxy with a high \hi mass in which the current star formation rate is low and star formation appears to have been largely quenched over the previous 1--2\,Gyr.

\subsection{Associated \mbox{H\,{\sc i}} absorption detection towards SDSS J113622+004854.}

\subsubsection{The \hi spectrum}
We also detected an associated \mbox{H\,{\sc i}} absorption line towards the galaxy SDSS J113622+004852 at redshift $z=0.563$. The full-band radio spectrum from SBID 13306 is shown in Fig. \ref{fig:J113622+004852_full_band_spectrum} and has an absorption line at 908.63 MHz. The \hi absorption redshift is consistent with the GAMA optical redshift, implying associated \hi absorption. An enlarged spectrum with rms noise of 2.98 mJy beam$^{-1}$\,channel$^{-1}$ is shown in Fig. \ref{fig:optical_radio_plot} and is best fitted with a single Gaussian component. The fitted parameters are listed in Table  \ref{tab:fitted_paras}. 

The estimated \mbox{H\,{\sc i}} column density for SDSS J113622+004852 is $1.35^{+0.10}_{-0.09}\times10^{21} $ atoms\,cm$^{-2}$ using Eq. \ref{equation:column_density} and assuming a covering factor of 1 and a spin temperature of 100\,K. 
\subsubsection{Optical properties} 
Fig. \ref{fig:optical_radio_plot} shows a $grz$ composite image of SDSS J113622+004852 from the DESI Legacy Imaging Surveys \citep{dey2019} with ASKAP radio contours overlaid.  
The accuracy of the SDSS optical photometry for this object is affected by the presence of a bright foreground star about 15\,arcsec away, and also visible in Fig. \ref{fig:optical_radio_plot}.

\begin{figure*}
    \includegraphics[width=17.5cm]{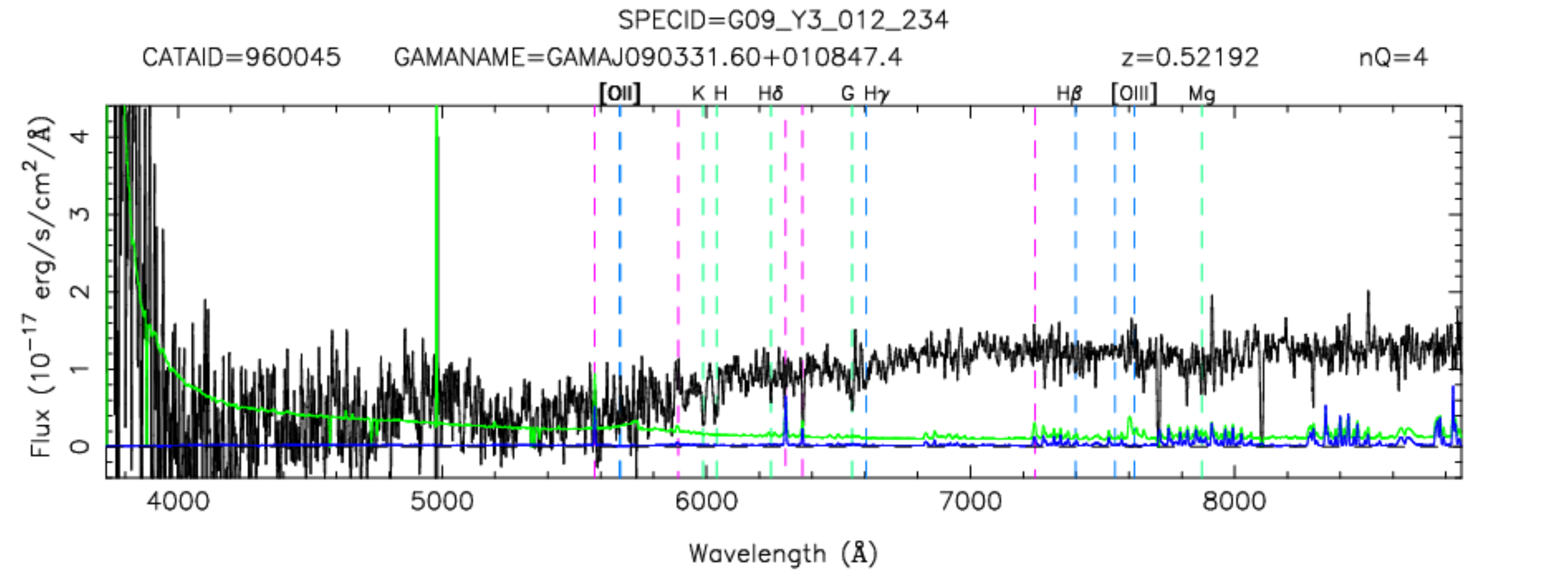}\\
    \includegraphics[width=17.5cm]{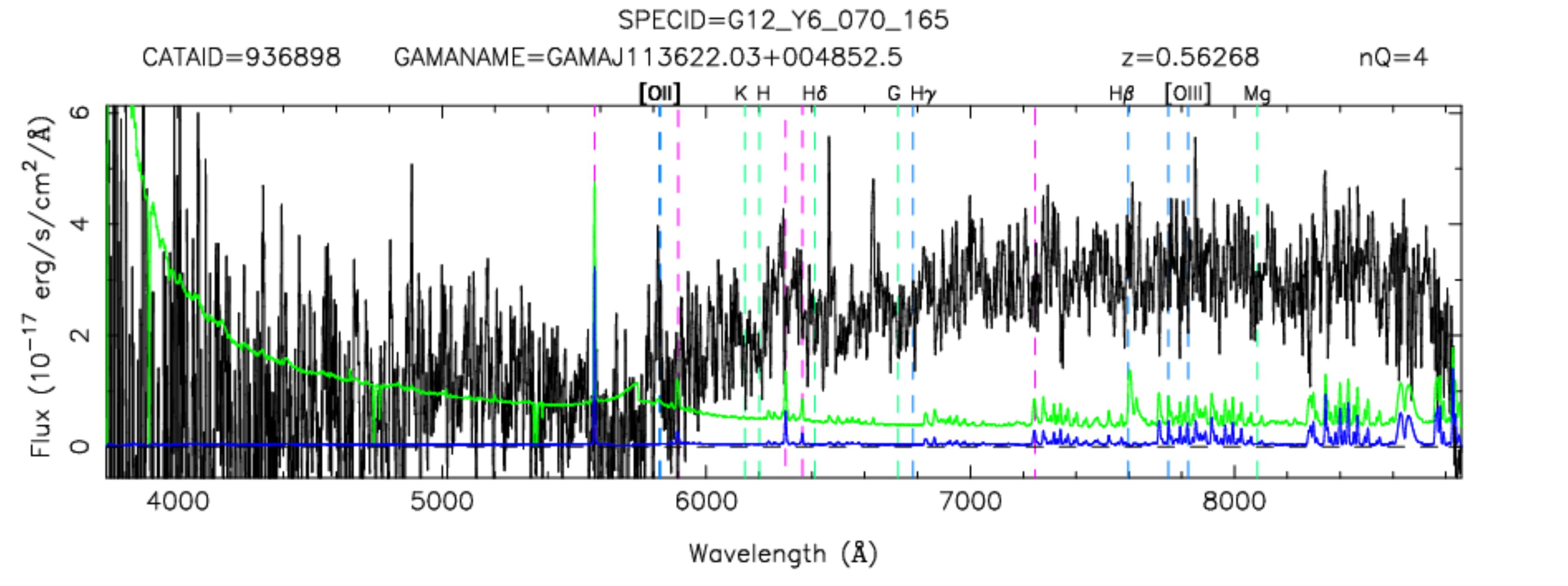}\\    
    \caption{GAMA optical spectra for SDSS J090331+010847 (upper) and SDSS J113622+004852 (lower). Although the S/N for both spectra is not high, the redshift measurements are reliable with $Q$ = 4. In each spectrum, the black line shows the source spectrum, the blue line is terrestrial night-sky emission and the green line a 1$\sigma$ error. The vertical lines indicate positions of spectral lines. }
    \label{fig:GAMA_spectra}
\end{figure*}

The GAMA optical spectrum of SDSS J113622+004852 gives a redshift of 0.563 \citep{baldry2018} with $Q=4$, corresponding to a systemic velocity of 168,900 km\,s$^{-1}$. The spectrum (see Fig. \ref{fig:GAMA_spectra}) shows weak [OII] emission but no other obvious emission lines, and was classified as a LERG by \cite{ching2017}. 

\begin{figure*}
    \includegraphics[width=9cm]{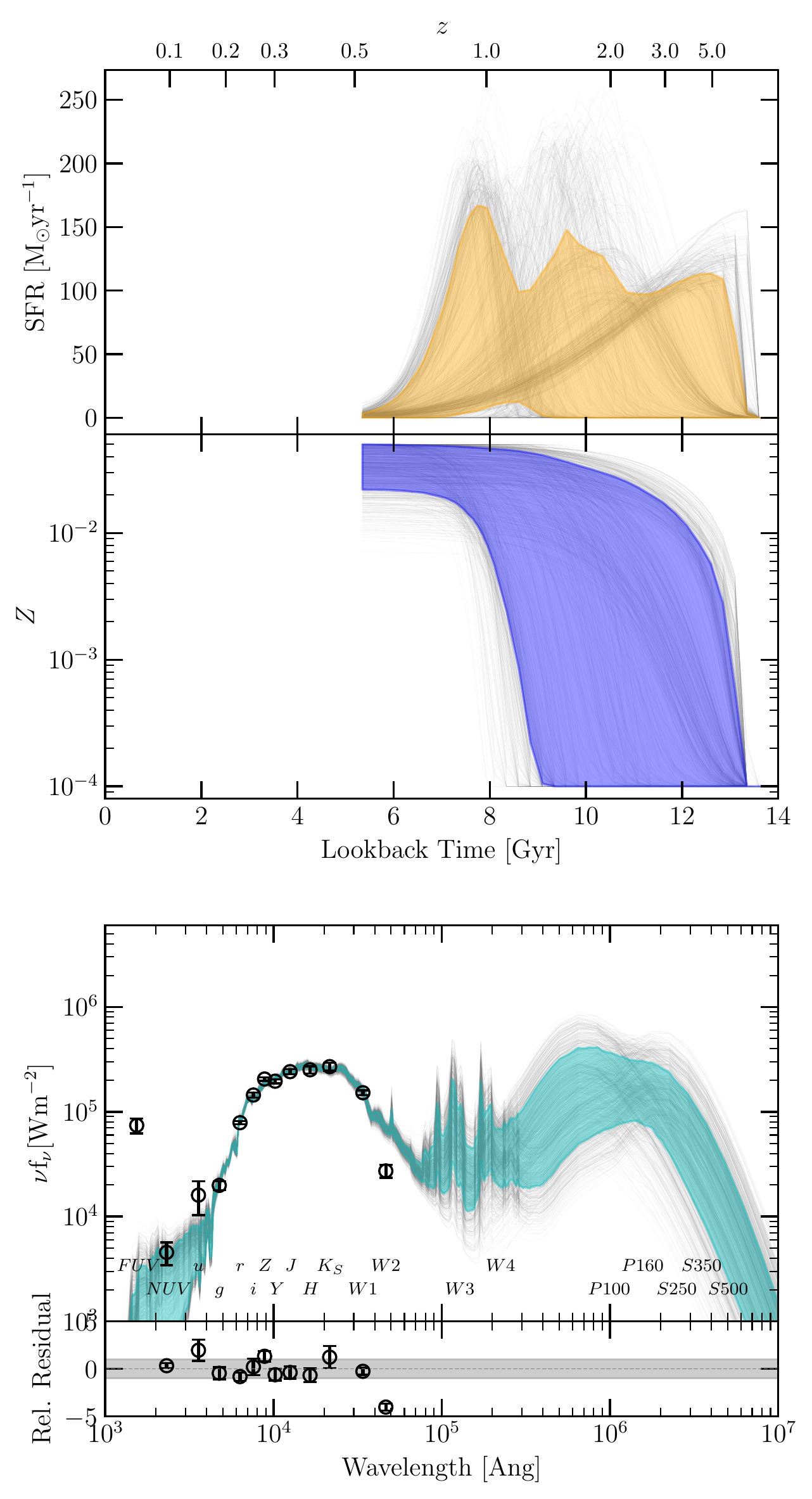}\includegraphics[width=9cm]{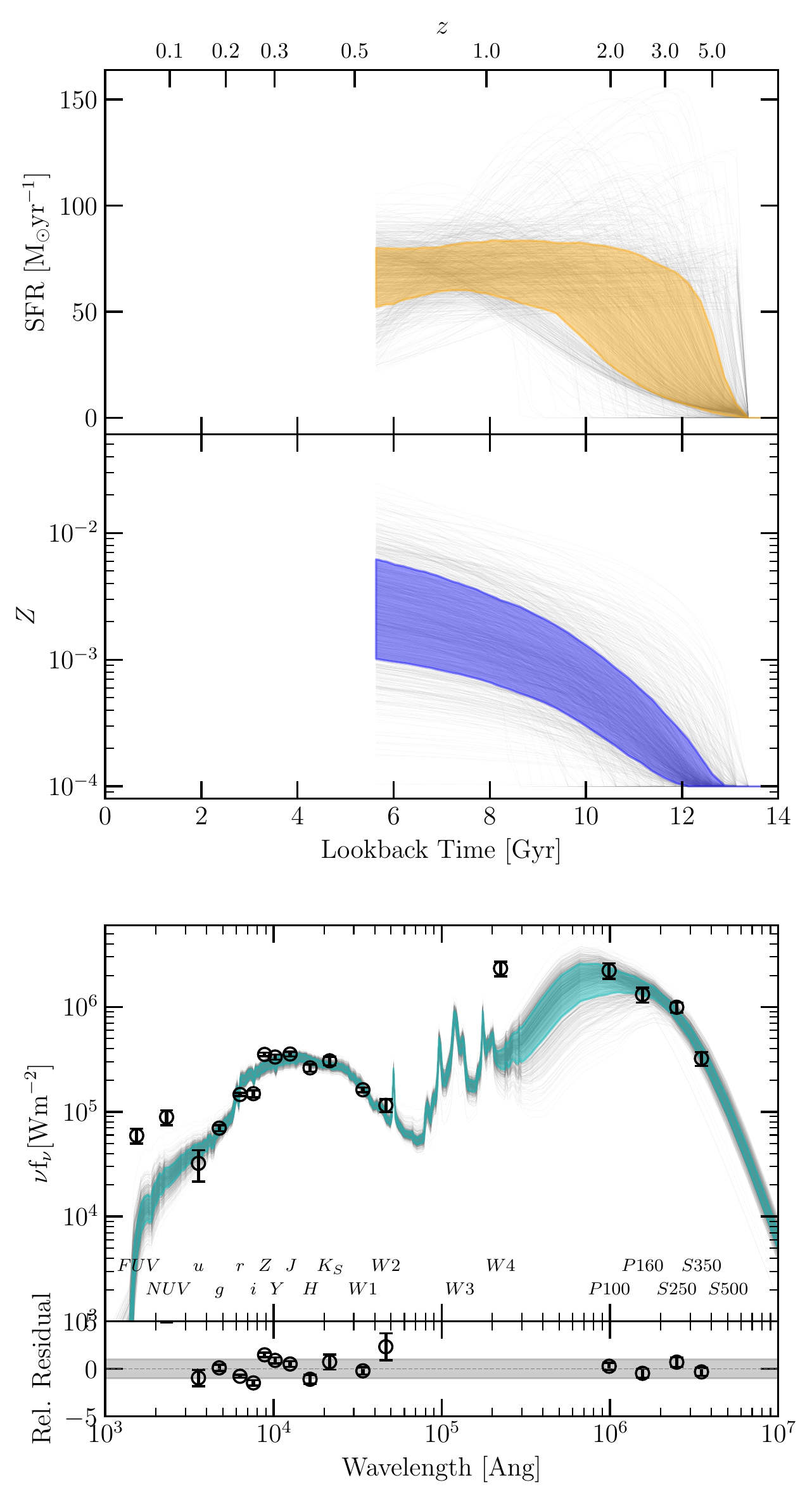}\\
    \caption{SED fitting results for SDSS J090331+010847 (left) and SDSS J113622+004852 (right) from the \prospect code \citep{robotham2020}. From top to bottom, they are star formation rate history, metallicity history, and UV-optical-infrared SED. The grey lines are 1000 iterations from the Markov Chain Monte Carlo (MCMC) posterior distribution with 1$\sigma$ range indicated by the coloured region in each panel. The photometric data in the bottom panel are from the GAMA panchromatic database \protect\citep{bellstedt2020}, constructed by extracting flux from each band image using \protect{\textsc {ProFound\ }} \protect\citep{robotham2018}. }
    \label{fig:optical_sed_fit}
\end{figure*}

The \prospect fit to the UV-optical-infrared SED for SDSS J113522+004852 is shown in Fig. \ref{fig:optical_sed_fit}. From the SED fitting, we derive a stellar mass of $2.33^{+0.47}_{-0.51}\times$ $10^{11} \rm M_{\odot}$, a metallicity $Z = 0.0025^{+0.0036}_{-0.0014}$, a dust mass of $2.19^{+1.35}_{-0.80}\times$ $10^{9} \rm M_{\odot}$, and a star formation rate of $68.65^{+11.75}_{-16.83}M_{\odot}\,yr^{-1}$. 
The metallicity derived for this galaxy is lower than would be expected from its stellar mass, since galaxies as massive as SDSS J113622+004854 generally have above-solar metallicity \citep[e.g.][]{bellstedt21}. We note however that metallicity information is a very subtle signature in the optical bands \citep{thorne21} and more accurate photometry is needed to determine whether this object is a genuine outlier from the mass-metallicity relation at intermediate redshift. 

\subsubsection{Radio properties}
We explored the radio SED of SDSS J113622+004852 
in the same way as for SDSS J090331+010857, by combining the data from our own observations with measurements from the literature. SDSS J113622+004852 is an unresolved source in FIRST, NVSS and VLASS, but Table \ref{table:target-sources} lists SDSS J113622+004852 as a two-component island in the ASKAP image (though with a separation smaller than the ASKAP beam) - raising the possibility that some extended low-frequency emission may also be present. 
 The fitted SED is shown in Fig. \ref{fig:radio_sed_fit} and fitted parameters are listed in Table  \ref{tab:radio_sed_fitted_pa}.

The radio spectrum of SDSS J113622+004852 peaks at 1279\,MHz in the observed frame, corresponding to 1999.1 MHz in the source rest frame, with an optically thick spectral index of $+0.36$ and optically thin index of $-1.78$. Like SDSS J090331+010847 therefore, SDSS J113622+004852 is also a peaked-spectrum  radio source. 

\subsubsection{Infrared properties}
The WISE W1, W2, and W3 magnitudes for SDSS J113622+004852 are 15.409$\pm$0.040, 14.839$\pm$0.068, and 9.799$\pm$0.055 respectivey. Its location in the WISE two-colour diagram is  shown by the cyan marker in Fig. \ref{fig:wise-color-color}. 

SDSS J113622+004852 has (W1-W2)$=0.57$, which is lower than the value of 0.8 often used to select AGN candidates  \citep{stern2012} and implies that the mid-IR emission comes mainly from the stellar host galaxy. This is consistent with the optical classification of LERG by \cite{ching2017}. 
However, the (W2-W3) value of 5.04 mag implies a star-formation rate at the high end of starburst galaxies, and this is consistent with the star formation rate (SFR) of 68.65 $\rm M_{\odot}\,yr^{-1}$ derived from the \prospect SED fit, which is much higher than the estimated SFR of 1.47 $\rm M_{\odot}\,yr^{-1}$ for SDSS J090331+010847.

\subsubsection{Summary}
The high \mbox{H\,{\sc i}} column density measured from \mbox{H\,{\sc i}} absorption and the properties derived from the UV-optical-infrared SED imply that SDSS J113622+004852 is a massive, gas-rich, and dust-obscured starburst galaxy. 
From the SED-derived dust mass, we estimate an \hi mass of at least $10^{11}$ M$_{\odot}$ for J113622+004852. 

In contrast to J090331+010847, the \prospect fit for SDSS J113622+004852 implies that it has sustained a high star-formation rate since at least $z\sim2$ and is continuing to form stars at the rate of over 50 $\rm M_{\odot}\,yr^{-1}$ with no sign of quenching. Its recent star formation history appears to be very different from that of J090331+010847, implying that the relationship between \hi and star formation at intermediate redshift is far from straightforward.

\begin{figure*}
    \includegraphics[width=18cm]{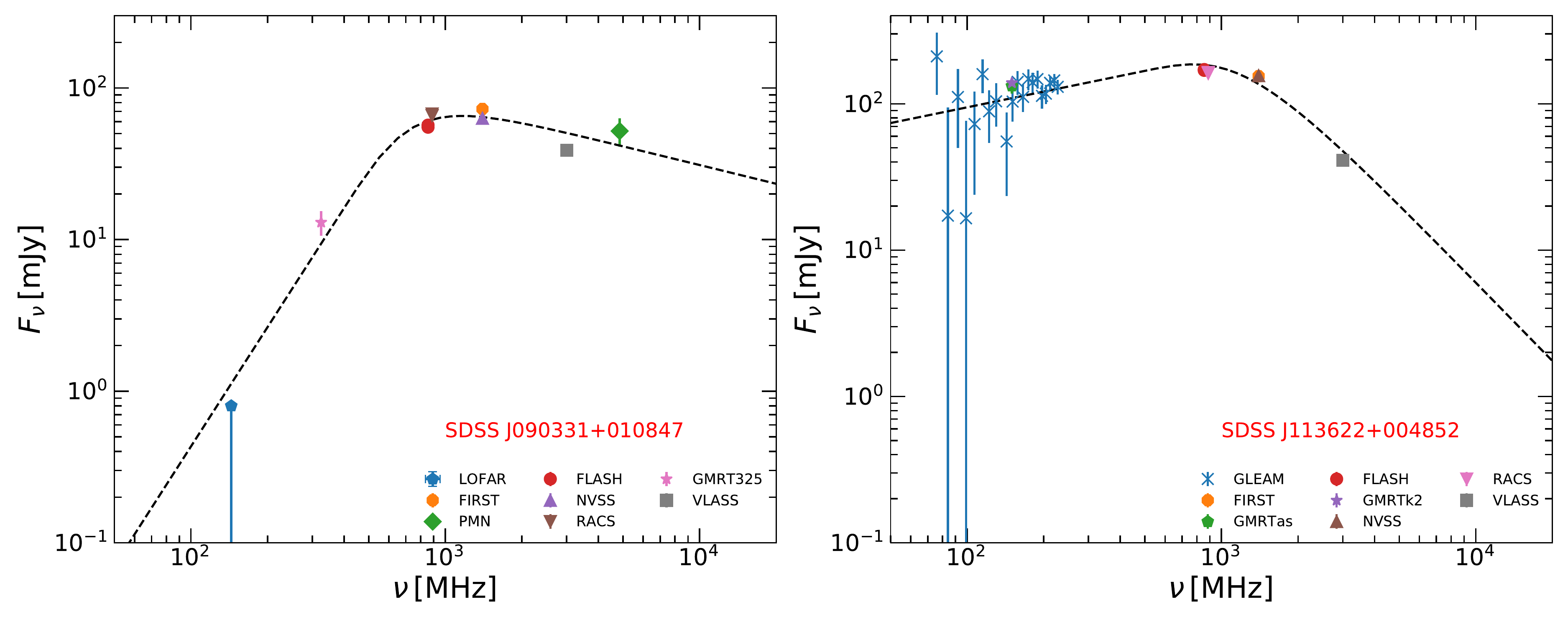}\\
    \caption{The fitted radio SEDs for SDSS J090331+010847 and SDSS J113622+004852. The frequency is given in the observed frame. References for the data: FLASH - this work; PMN - \protect\cite{griffith1995}; VLASS - \protect\cite{gordon2020}; NVSS - \protect\citep{condon1998}; FIRST- \protect\citep{white1997}; GMRT325 - \protect\citep{mauch2013}; RACS - \protect\citep{hale2021}; GMRTas - \protect\citep{intema2017}; GLEAM - \protect\citep{hurley2017}; GMRTk2 - \protect\citep{tingay2016}. The LOFAR upper limit at 144\,MHz was provided by J.R. Callingham and T.W. Shimwell (private communication). }
    \label{fig:radio_sed_fit}
\end{figure*}

\begin{figure*}
    \includegraphics[width=16cm]{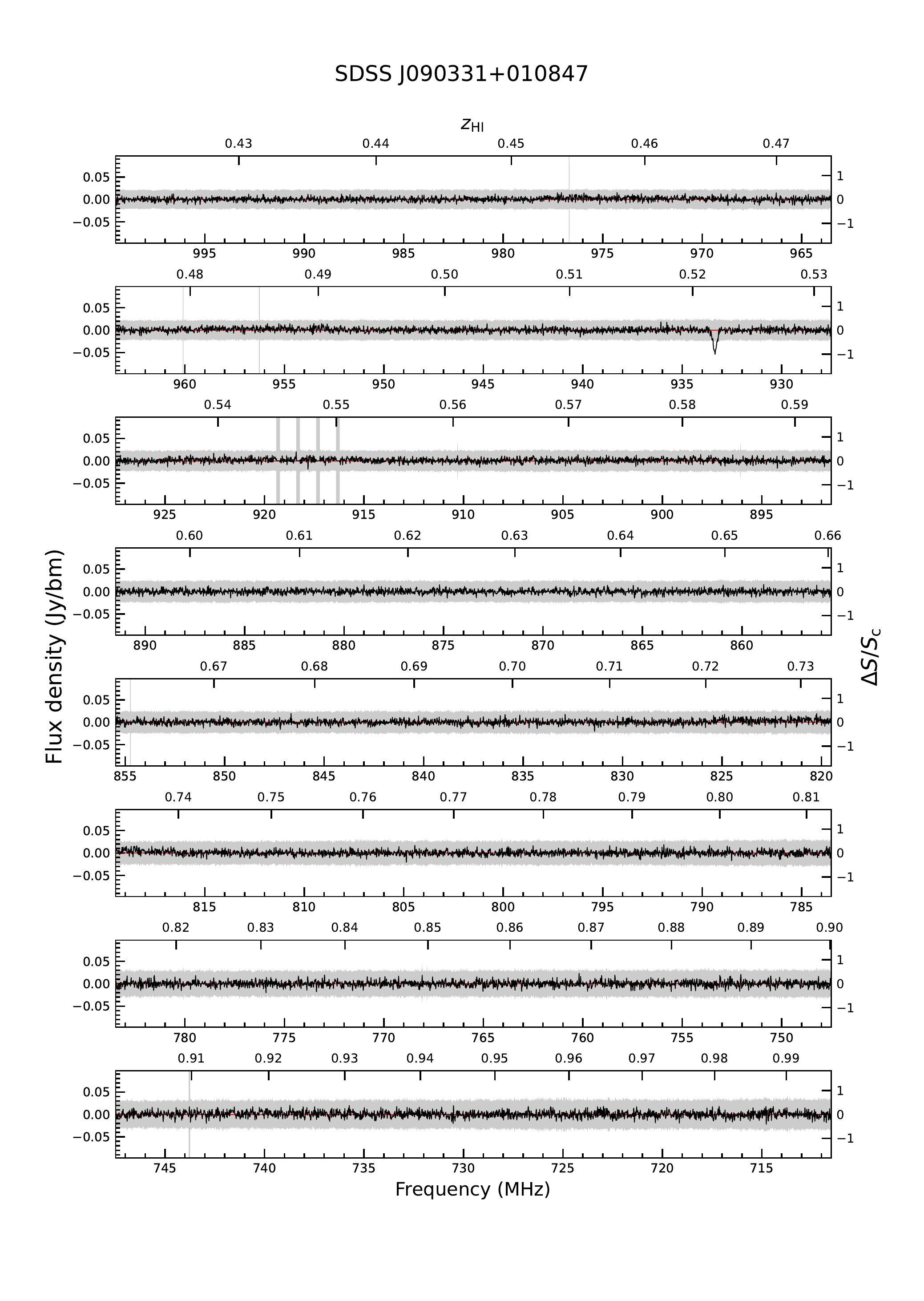}
    \caption{The full-band spectrum towards SDSS J090331+010847, from 711.5 MHz to 999.5 MHz. The black line is the continuum subtracted flux density (left Y-axis), or the fraction of the continuum flux density (right Y-axis), as a function of frequency (lower X-axis) or redshift (upper X-axis). The grey region denotes 5 times the rms noise. An absorption line is visible at 933.352 MHz, which is consistent with the GAMA spectroscopic redshift of SDSS J090331+010847. }
    \label{fig:J090331+010847_full_band_spectrum}
\end{figure*}

\begin{figure*}
    \includegraphics[width=16cm]{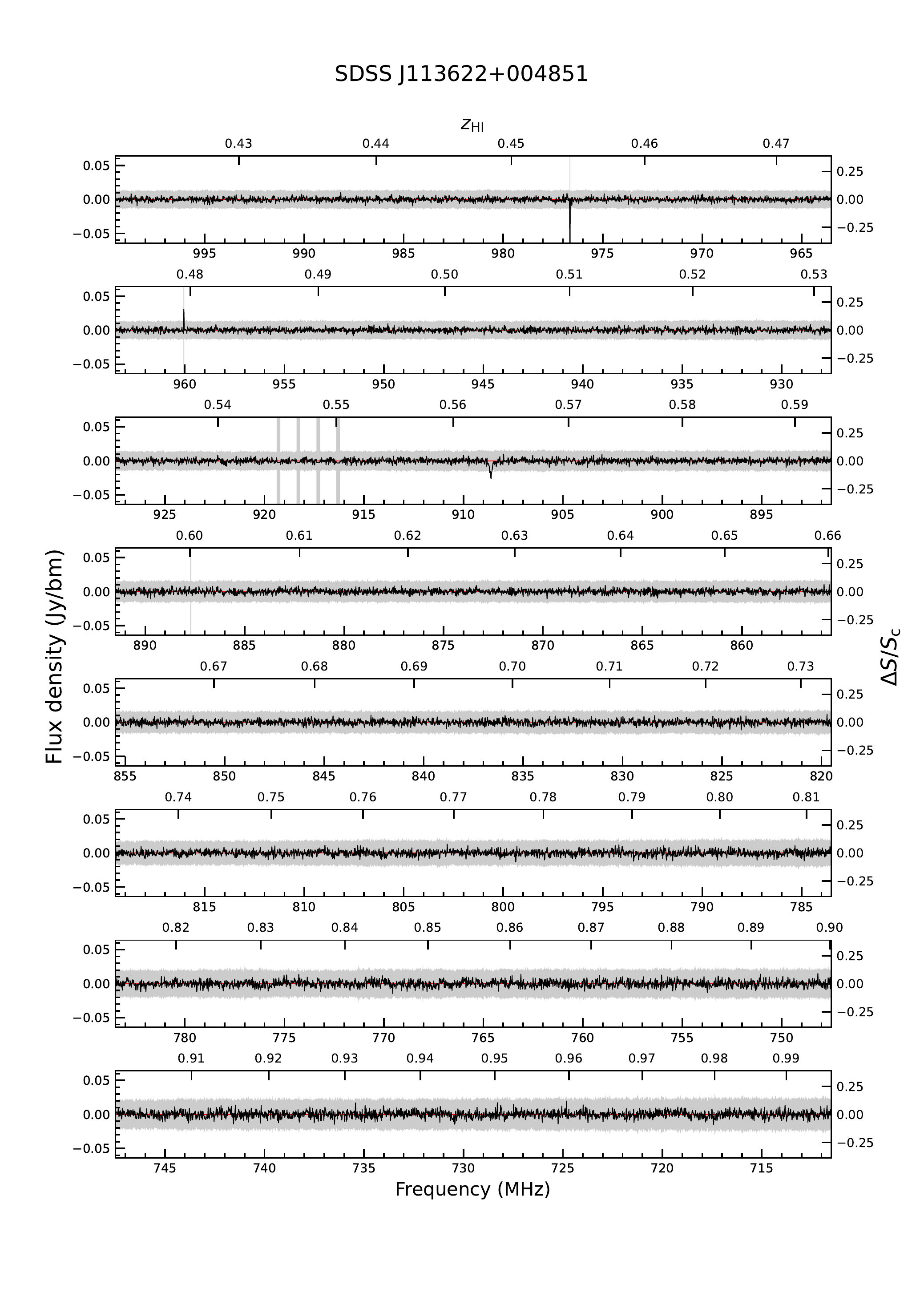}
    \caption{The same as Fig. \ref{fig:J090331+010847_full_band_spectrum} but for SDSS J113622+004852. }
    \label{fig:J113622+004852_full_band_spectrum}
\end{figure*}

\begin{figure*}
    \includegraphics[width=6.5cm]{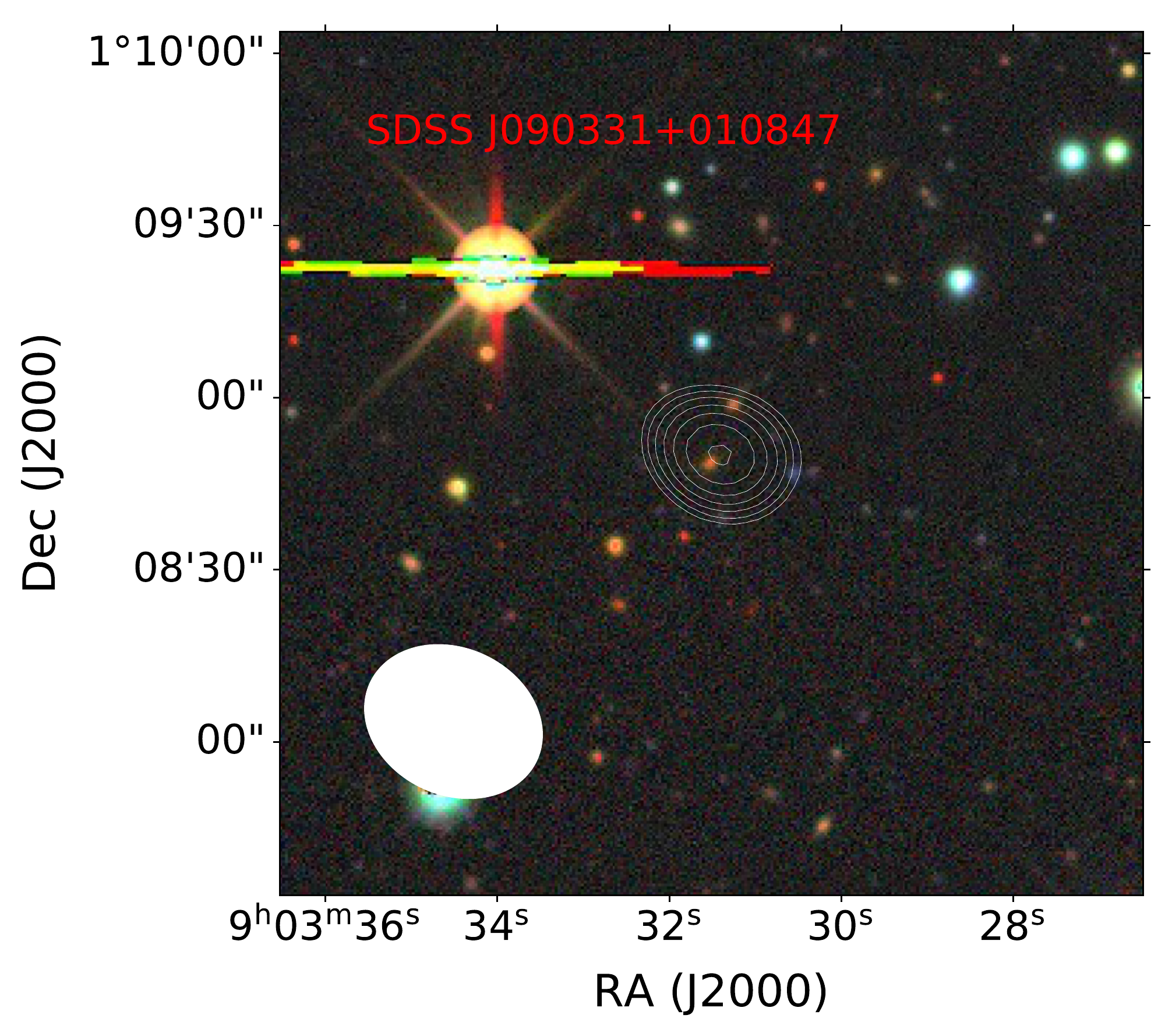}\includegraphics[width=9cm]{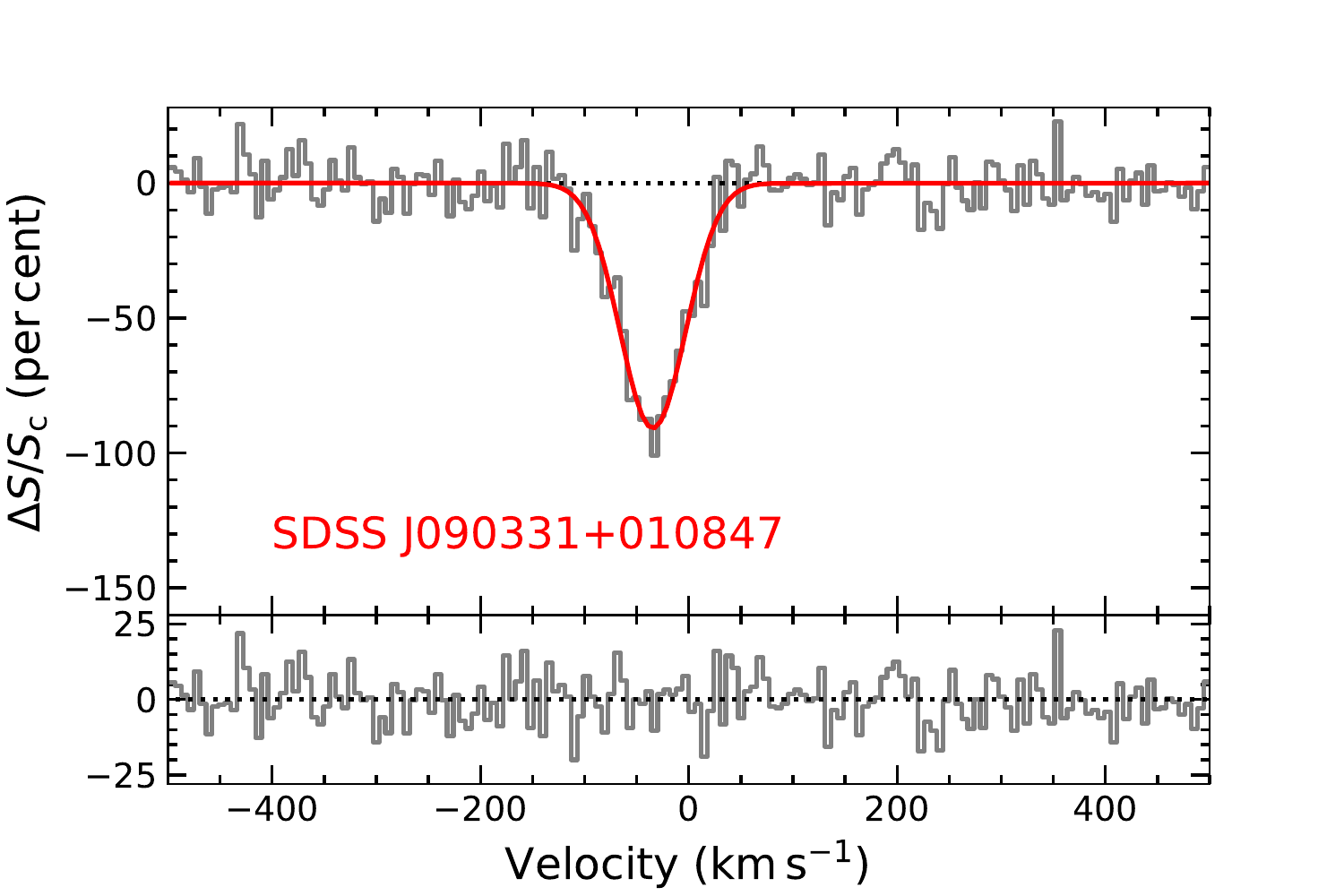}\\
        \includegraphics[width=6.5cm]{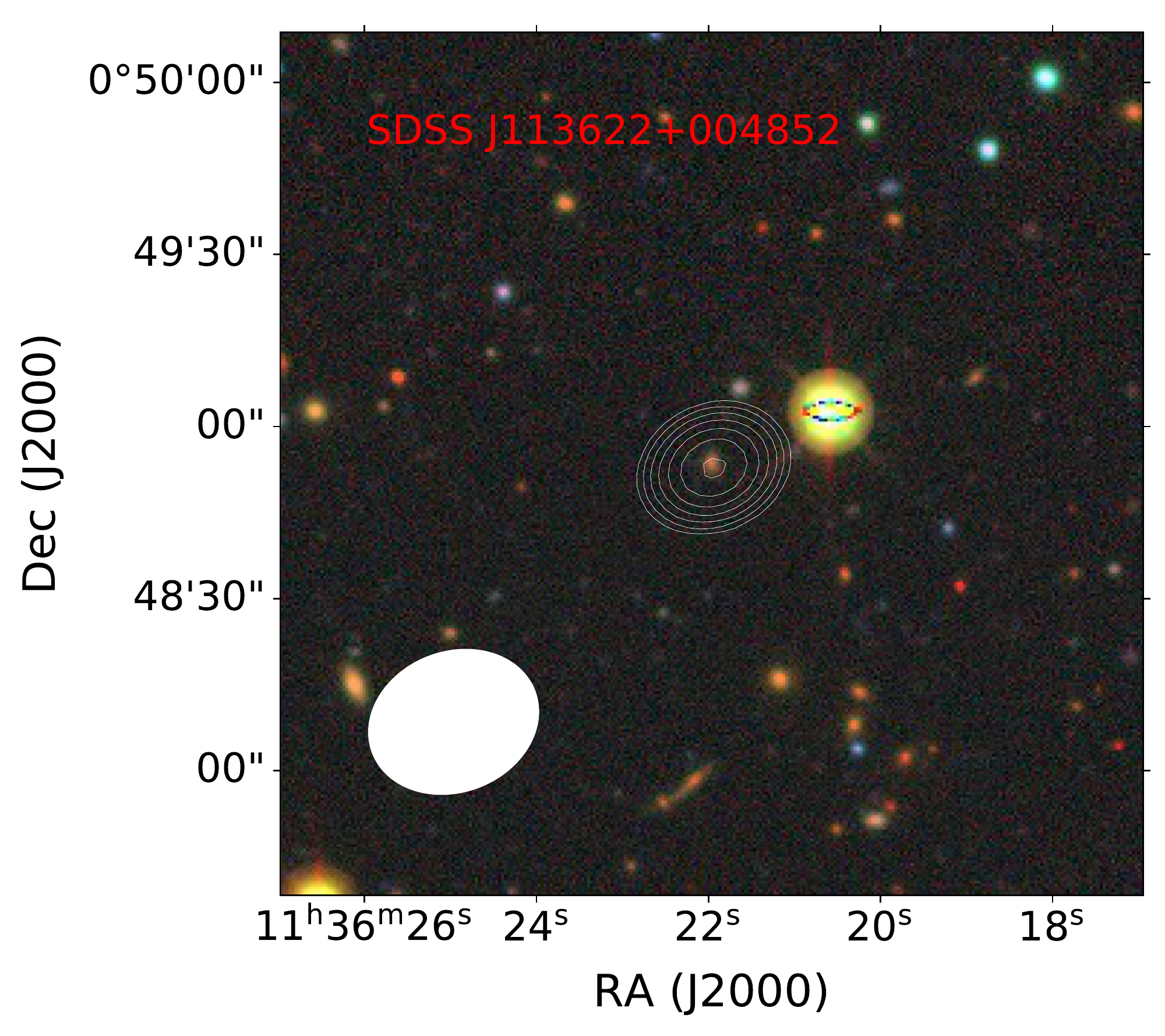}\includegraphics[width=9cm]{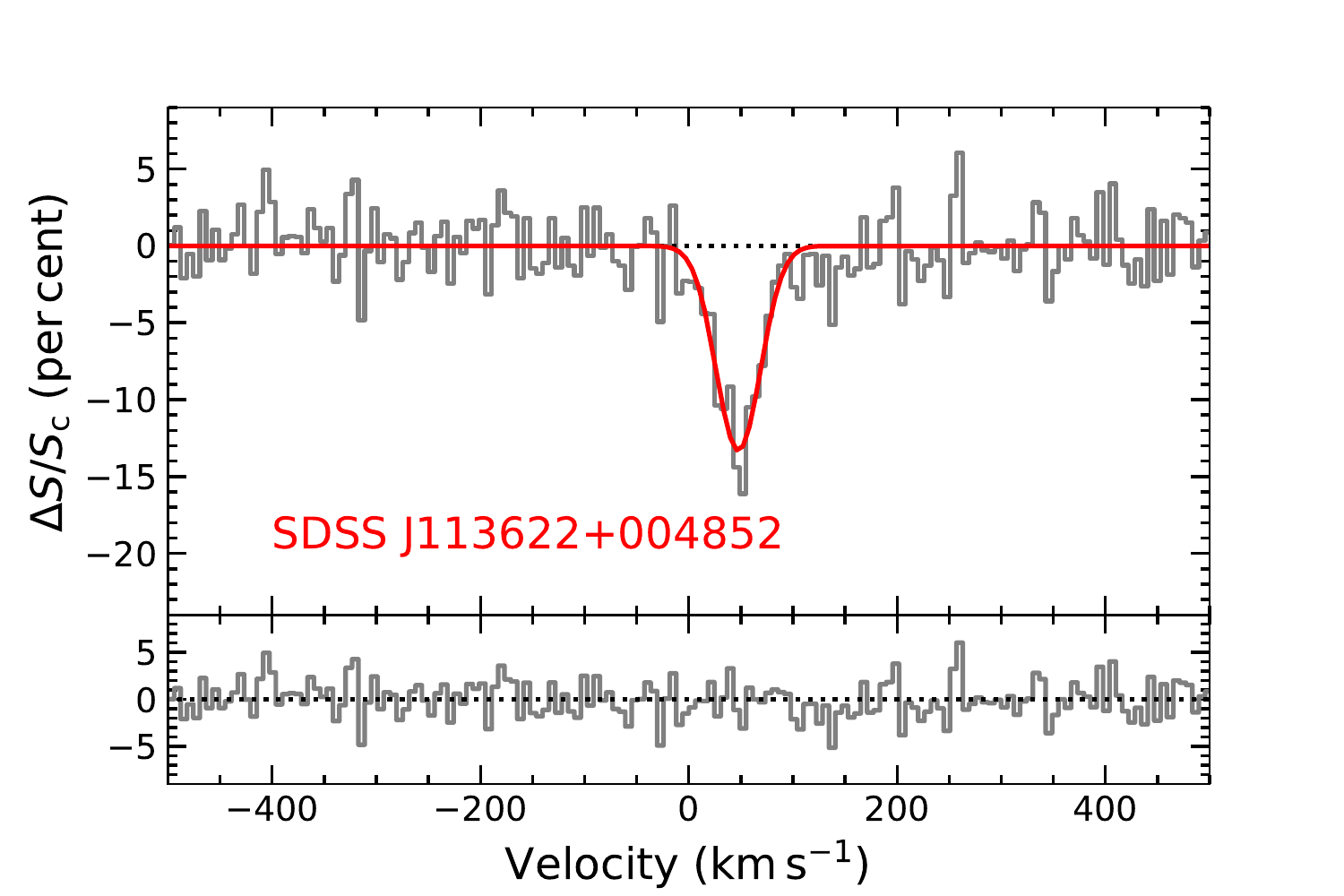}
    \caption{ Upper-left: the $grz$ composite images of SDSS J1090331+010847 from DESI Legacy Imaging Surveys \citep{dey2019} with radio contours at 855.5 MHz from our ASKAP continuum observation overlaid. The radio contours start at 6.28 mJy\,beam$^{-1}$ and increase by factors of $\sqrt{2}$. The ASKAP beam is plotted at left-lower corner. Upper-right: the zoom-in \mbox{H\,{\sc i}} absorption spectrum around the systemic velocity of SDSS J1090331+010847. The grey line is continuum subtracted spectral data from observation with SBID 13283 and showed in fraction of peak continuum flux density. The red line is fitted Gaussian profile and the residual is plotted in its lower part. The lower plots are the same as upper plots but for the source of SDSS J113622+004852 which has a start radio contour of 19.4 mJy\,beam$^{-1}$. The spectrum shown is from observation SBID 13306.}
    \label{fig:optical_radio_plot}
\end{figure*}

\subsection{Upper limits for \mbox{H\,{\sc i}} non-detections}
For the 324 sources that were not detected in \mbox{H\,{\sc i}} absorption, we now estimate a 3$\sigma$ upper limit 
for the \hi column density. 
 
For a Gaussian absorption line, the integrated optical depth is $\int \tau(V) d{V}$ = $\sqrt{\pi/\ln 2}\times \tau_{\rm pk}\times {\rm FWHM}/2 $, where $\tau_{\rm pk}$ is the peak optical depth. 
We can combine this relation with  Eq. \ref{equation:column_density} to calculate an upper limit in \nhi\ for each source assuming a Gaussian line profile:

\begin{equation}\label{equation:column_density_optically_thin}
\begin{split}
N_{\rm HI,lim} = 1.82\times10^{20}\times T_{s} \times \sqrt{\pi/\ln 2}\times \tau_{\rm pk}\times {\rm FWHM}/2 \\= 2.32\times10^{20}\times T_{s}\times \ln\left(\frac{C_{f}S_{\rm cont}}{C_{f}S_{\rm cont}-3\sigma}\right)
\end{split}
\end{equation}

where $\sigma$ is the rms spectral noise listed in Table \ref{table:target-sources-infor}. For 49 of our faintest radio targets the 3$\sigma$ noise is larger than their continuum flux density, so we only calculate a 3$\sigma$ upper limit for the remaining 275 sources. 
We adopt values of $T_{s}$ = 100\,K and  $C_{f}$ = 1, and assume FWHM = 120 km\,s for an associated \hi absorption line \citep{allison2022}.  
The calculated 3$\sigma$ peak optical depth upper limit and the corresponding \mbox{H\,{\sc i}} column density upper limit for each source are listed in columns 20 and 21 of Table \ref{table:target-sources-infor}.

\begin{figure}
    \includegraphics[width=8cm]{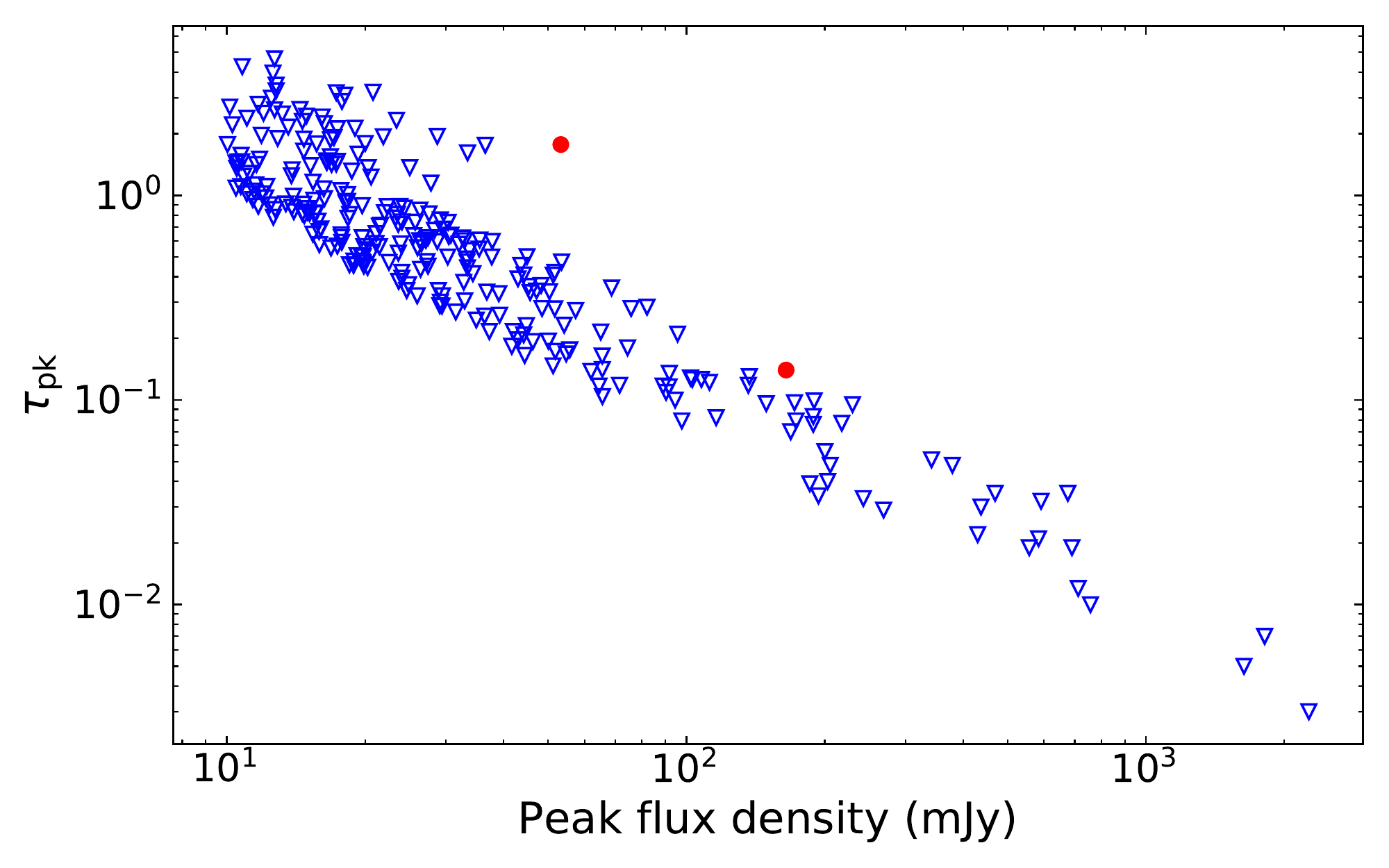} \\
    \caption{The blue triangles denotes the 3$\sigma$ upper limit of peak optical depth as a function of peak flux density for 275 non detections. The two red circles denotes the peak optical depth of two detected \mbox{H\,{\sc i}} absorbers.}
    \label{fig:tau_flux}
\end{figure}

Fig. \ref{fig:tau_flux} shows the 3$\sigma$ upper limit in peak optical depth as a function of peak flux density for all the sources in our target sample.  For most of these objects, we have limited sensitivity to lines with low \hi optical depth. For sources with peak flux density above about 100\,mJy we can make 3$\sigma$ detections of lines with $ \tau_{\rm pk}\sim 0.1$.  This is comparable to the detection limit of $ \tau_{\rm pk}\sim0.08$ for the \cite{maccagni2017} sample, but less sensitive than the limit of $ \tau_{\rm pk}\sim0.01$ achieved by \cite{vermeulen2003} for a sample of 57 sources. 

Very strong lines like that seen in SDSS J090331+010847 with $ \tau_{\rm pk}=1.77$ can however be detected against much fainter sources, and the inclusion of large numbers of relatively faint continuum sources in a single observation is one advantage of a wide-field absorption survey with ASKAP. 

\begin{table*}
\caption{The fitted SED parameters for SDSS J090331+010847 and SDSS J113622+004852. The errors are determined from 1000 fittings.}
\centering
\begin{tabular}{lllll}
\hline
Name & ${\rm S_{max}}$ (mJy) &$\nu_{\rm max}$ (MHz) & $\alpha_{\rm thick}$ &$\alpha_{\rm thin}$   \\
\relax
[1] & [2] & [3] & [4] & [5]\\
\hline
SDSS J090331+010847 & 56 $\pm$7 &  768$\pm$152 & 2.61$\pm$0.68 & -0.41$\pm$0.22  \\
SDSS J113622+004852 & 149$\pm$35 &  1279$\pm$577 & 0.36$\pm$0.81 & -1.78$\pm$1.06\\
 
\hline
\end{tabular}
\label{tab:radio_sed_fitted_pa}
\end{table*}

\section{The detection rate for associated \mbox{H\,{\sc i}} absorption}\label{HI_rate}

\subsection{Detection rate from this survey}
We found two associated \mbox{H\,{\sc i}} absorption detections from a sample of 326 targets with $S_{855.5} \geq 10$ mJy. 
At first glance, this detection rate ($<1$\%) is much lower than that in found in previous studies of low redshift sources \citep[e.g.][]{van1989,emonts2010,chandola2011,glowacki2017}. However, the samples in several of these studies were pre-selected to be compact radio sources, dust-obscured galaxies, or very bright galaxies.  

A useful comparison for our ASKAP sample is the representative study of a sample of 248 low-redshift ($0.02<z<0.25$) sources observed with the Westerbork radio telescope (WSRT). The WSRT observations had integration times of up to six hours and achieved a median rms noise level of 0.81\,mJy\,beam$^{-1}$ 
\citep{gereb2014,gereb2015,maccagni2017}. The detection rate for associated \hi absorption in the WSRT sample was $27\pm5.5$\% for sources with a 1.4\,GHz core flux density larger than 30 mJy.  

In contrast to the WSRT study, FLASH is a relatively shallow survey with a typical rms spectral-line noise of $\sim$ 4.6 mJy in a 2 hour observation (see Table \ref{table:obs_summary}). 
To compare our results with the lower-redshift WSRT sample, 
we therefore compare the detection rate of \mbox{H\,{\sc i}} 21-cm absorption with optical depth $\tau>0.1$. In our observations, 35 of the 326 sources are bright enough to allow us to detect $\tau_{\rm pk}$ > 0.1 \mbox{H\,{\sc i}} gas with at least 3$\sigma$ significance and we detected one of these sources,  SDSS J113622+004852. 
(Our another associated absorber SDSS J090331+010847 is not one of these 35 sources due to its relatively low peak flux density). 
Our detection rate for sources with $\tau_{\rm pk}$ > 0.1, defined as DR$(\tau_{\rm pk}>0.1)$, is $2.9^{+9.7}_{-2.6}$\% (where the  quoted errors on the detection rates calculated in this paper are the 95\% confidence level for a binomial distribution) \citep{brown2001}.

\begin{figure}
    \includegraphics[width=8cm]{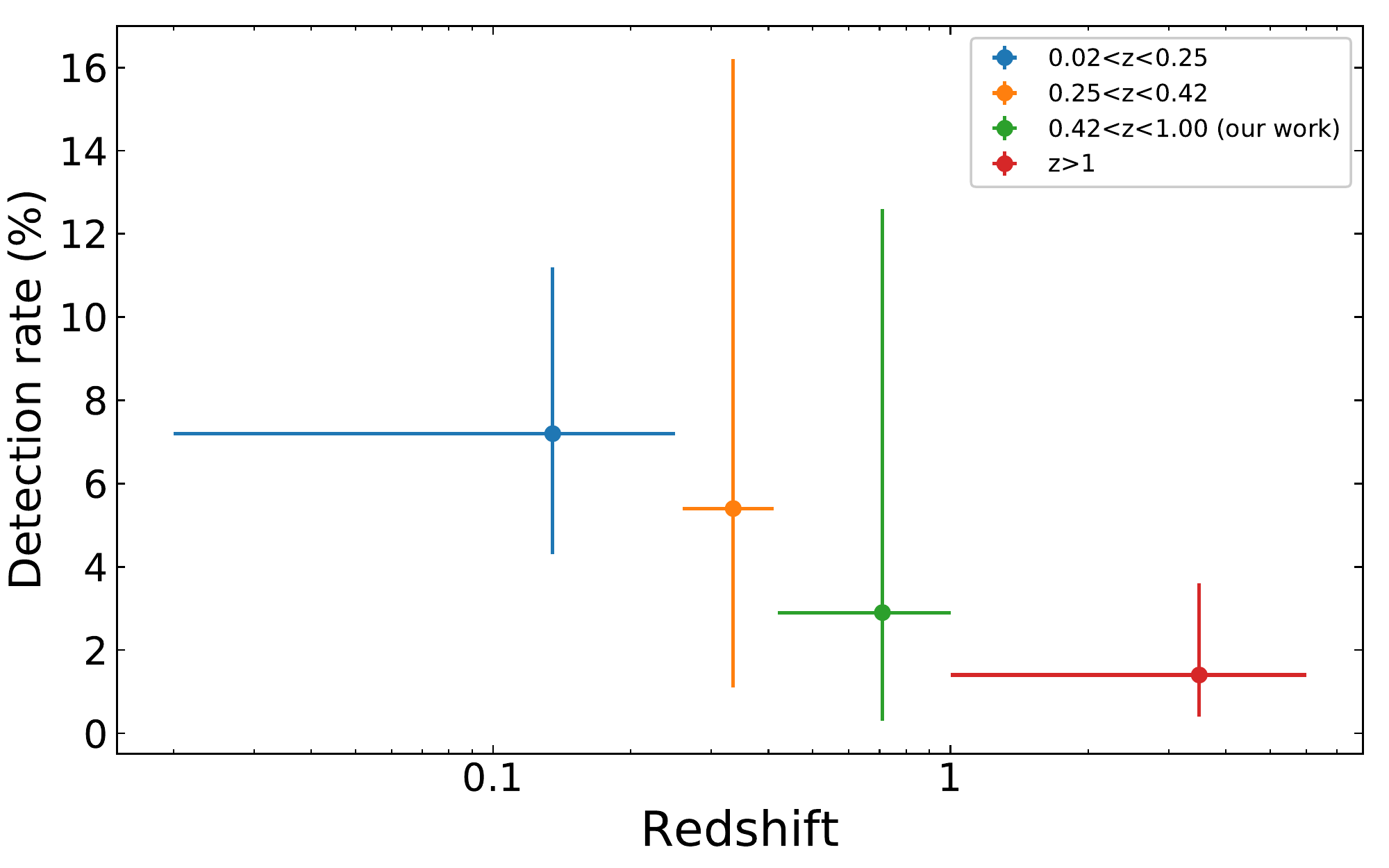}\\
    \caption{The detection rate for \mbox{H\,{\sc i}} gas with optical depth greater than 0.1 and with at least 3$\sigma$ significance across various redshift ranges. The blue, orange, green, and red markers indicate the detection rate from redshift 0.02 to 0.25 \citep{maccagni2017}, 0.25 to 0.42 (\citealt{aditya2018b,grasha2019,murthy2021}), 0.42 to 1.00 (our work), and > 1 (\citealt{aditya2016,aditya2017,aditya2018a,aditya2018b,aditya2021,carilli2007,Chowdhury2020,curran2006,curran2008,curran2011,curran2013b,curran2013a,curran2016,curran2017,dewaard1985,Dutta2020,grasha2019,gupta2006,gupta2021,ishwara2003,kanekar2009,mhaskey2020,rottgering1999,uson1991,yan2016}), respectively. }
    \label{fig:detection_rate}
\end{figure}

\subsection{Comparison with other surveys at different redshifts}
To make a fair comparison with previous surveys at other redshifts, we choose to compare the occurrence of \mbox{H\,{\sc i}} gas with $\tau_{\rm pk}$ > 0.1 for the reasons explained above. 

In the redshift range $0.02<z<0.25$, we used the data from  \cite{maccagni2017}. There are 222 sources in this sample for which it was possible to detect lines with $\tau_{\rm pk}>0.1$ with a significance greater than 3$\sigma$ and 16 of these sources were detected, giving a  detection rate, DR$(\tau_{\rm pk}>0.1)$ of $7.2^{+2.9}_{-4.0}$\%.

In the redshift range $0.25<z<0.42$, we consider several literature studies. \cite{murthy2021} used Karl G. Jansky Very Large Array (JVLA) to observe 26 radio-loud AGN with FIRST peak flux densities of 400--800 mJy\,beam$^{-1}$. 
\cite{aditya2018b} compiled observations of  Caltech-Jodrell Bank Flat-spectrum (CJF) sources at redshift $0.01<z<3.6$ that had been searched for \mbox{H\,{\sc i}}, and \cite{grasha2019}  compiled a sample of 145 compact and flat-spectrum radio sources at $0<z<4$.  to do a comparison. Between these three studies, there are 37 sources in the redshift range $0.25<z<0.42$ that are suitable for detecting $\tau_{\rm pk}$ > 0.1 with a significance above 3$\sigma$. Two of these sources were detected with $\tau_{\rm pk}$ > 0.1, giving a detection rate, DR$(\tau_{\rm pk}>0.1)$, of $5.4^{+10.8}_{-4.3}$\% in this redshift range. 

In the redshift range $0.42<1.0$ there are a few previous studies (e.g. \citealt{aditya2018b,aditya2019,murthy2022}), but we chose to us our own data to minimize any selection bias. As noted above, the detection rate here is $2.9^{+9.7}_{-2.6}$\%.

For high-redshift searches at $z>1$, we explored almost all the available literature  (\citealt{aditya2016,aditya2017,aditya2018a,aditya2018b,aditya2021,carilli2007,Chowdhury2020,curran2006,curran2008,curran2011,curran2013b,curran2013a,curran2016,curran2017,dewaard1985,Dutta2020,grasha2019,gupta2006,gupta2021,ishwara2003,kanekar2009,mhaskey2020,rottgering1999,uson1991,yan2016}) and found three detections with $\tau_{\rm pk}>0.1$ out of 221 sources for which 3$\sigma$ detections could have been made at this level. This gives a detection rate DR$(\tau_{\rm pk}>0.1)$ of  $1.4^{+2.2}_{-1.0}$\%.

We plot the detection rates across these different redshift ranges in Fig. \ref{fig:detection_rate}. 
Although the error bars are large, we can see a general trend for the \hi detection rate to decrease with increasing redshift. This is consistent with the findings of other studies, \citep[e.g.][]{aditya2016,aditya2018b}.

Previous studies \citep[e.g.][]{maccagni2017} found a higher detection rate for \hi absorption in compact (and gas-rich) radio galaxies compared to extended radio galaxies.
Here, we tentatively compare our results with the WSRT survey by  \cite{maccagni2017} since both the FLASH and WSRT surveys are large and relatively unbiased with respect to source morphology. 

\cite{maccagni2017} divided their sources into three classes based on their WISE colours: `dust-poor' (W1–W2 < 0.5 and W2–W3 < 1.6), `12-$\mu$m-bright' (W1–W2 < 0.5 and 1.6 < W2–W3 < 3.4), and `4.6-$\mu$m-bright' (W1–W2 > 0.5), and most of their sources were classified as dust-poor and 12-$\mu$m-bright. In the FLASH survey, most of our targets with reliable W1, W2, and W3 measurements have 0 < W1–W2 < 0.5 and 2 < W2–W3 < 4 so would be classed as `12-$\mu$m-bright' in this scheme (see Fig. \ref{fig:wise-color-color}). 

The classification of our detected source SDSS J090331+010847 is uncertain in the \cite{maccagni2017} scheme because of the unrestrictive upper limit in W3, whereas SDSS J113622+004852 can be classified as a `4.6-$\mu$m-bright' source -- allowing us to compare the detection rates for this class. 
In the WSRT survey, there are 44 4.6-$\mu$m-bright sources (of the 49 4.6-$\mu$m-bright sources in the table A1 of \cite{maccagni2017}) suitable for detecting $\tau_{\rm pk}>0.1$ with a significance above 3$\sigma$. Six sources were detected with these requirements, giving a detection rate of $13.6^{+12.3}_{-7.7}$\%. 
In our observations, we obtained one detection out of 28, giving a detection rate of $3.6^{+11.9}_{-3.2}$\% which is lower than that in the WSRT survey. 

Similarly, the corresponding detection rate for compact radio sources in the WSRT sample,  $9.7^{+6.5}_{-4.5}$\%, is higher than our rate of of $3.6^{+11.9}_{-3.2}$\%, with one detection from 28 FLASH compact sources that were bright enough for a line to be detectable at the level of $\tau_{\rm pk} = 0.1$. 
In both cases however, the small number of associated \hi detections in our sample means that the differences between the two samples are suggestive rather than statistically significant. 

It would be interesting to know whether the \hi detection rate for the various classes of radio galaxies evolves with redshift, but associated \hi detections with $\tau_{\rm pk}>0.1$ at  redshift $z > 0.25$ are quite rare and more detections are needed to allow a detailed comparison, which is outside the scope of the current paper. 

\subsection{Possible reasons for a varying \hi detection rate} 
There is growing evidence from the literature that the detection rate for associated \hi absorption decreases at higher redshift, and we briefly discuss several possible reasons for this. 
\subsubsection{Redshift evolution of the \hi content of radio-loud AGN} 
We can rule out a scenario in which the decline in \hi detection rate with redshift occurs because there is less \hi gas in higher redshift galaxies. Ly$\alpha$ absorption and \hi emission-line stacking studies have shown that this is not the case, and that the overall cosmic \hi mass density increases at higher redshift (e.g. see Fig. 14 in \citealt{rhee2018}). 

\subsubsection{UV luminosity effects and ionisation}
Many studies for associated \mbox{H\,{\sc i}} absorption in high redshift AGN pre-selected optically-bright sources that are luminous in the rest-frame UV band. 
\cite{curran2008} and \cite{curran2012} pointed out that if the UV luminosity of an AGN exceeds some critical value, all the neutral hydrogen in that galaxy could be ionised and no longer observable as \hi.  This hypothesis has been confirmed and applied widely (e.g. \citealt{curran2011,curran2016,curran2017,aditya2016, grasha2019}). 

As can be seen from Table \ref{table:target-sources-infor} and Fig. \ref{fig:sample_properties} however, the ionising photon rate distribution of our targets peaks around $10^{52}$\,s$^{-1}$ and only a few objects have an ionising photo rate above the critical value of  $1.7\times10^{56}$\,s$^{-1}$. Thus it appears that the UV luminosity effect cannot account for the low detection rate in our sample.     

\subsubsection{Effects related to radio luminosity}
Because the sensitivity of an \hi absorption search depends on the flux density of the radio source used as a probe, the sources searched at higher redshift are likely to have higher radio luminosity. 
However, the WSRT search of 248 compact and extended sources did not find any significant difference in detection rate across a range of  radio luminosity from P$_{1.4\rm GHz}=10^{22.5}\,\rm W\,Hz^{-1}$ to P$_{1.4 \rm GHz}=10^{26.2}$\,W\,Hz$^{-1}$ (\citealt{gereb2014,maccagni2017}).

\cite{aditya2016} suggested a lower detection rate towards higher radio luminosity ($P_{1.4\rm GHz}>10^{27.3}\,\rm W\,Hz^{-1}$) by analysing 52 flat-spectrum sources, but a study by \cite{curran2018} that compiled all known $z>0.0021$ \mbox{H\,{\sc i}} absorption detections found that the detection rate did not vary significantly with radio luminosity. We therefore conclude that radio luminosity effects are unlikely to account for the relatively low detection rate in our sample. 

\subsubsection{Changes in \hi spin temperature}
The gas spin temperature $T_{\rm spin}$ can affect the detection rate for an \hi absorption survey because for a fixed \hi column density the observed 21\,cm  optical depth $\tau$ will decrease as $T_{\rm spin}$ increases. 
 The spin temperature was often assumed to be 100\,K in previous studies when there is no direct measurement. 
 
 For distant galaxies, combining Ly$\alpha$ absorption and \mbox{H\,{\sc i}} absorption has been used to successfully measure the spin temperature. By analysing 37 DLA systems observed in \mbox{H\,{\sc i}} absorption, \cite{kanekar2014}
 found a statistically significant difference between the T$_{\rm spin}$ distributions in their high-redshift ($z>2.4$) and low-redshift ($z<2.4$) sub-samples, with higher average values of $T_{\rm spin}$ at high redshifts. Additionally, \cite{allison2021} used statistical methods to measure the harmonic mean spin temperature at redshifts between $z=0.37$ and 1.0 and found evidence that the typical \hi gas at these redshifts  may be colder than the Milky Way interstellar medium. 

The \cite{kanekar2014} and \cite{allison2021} spin temperature studies were both carried out on samples of intervening \hi absorbers, rather than the associated systems studied in this paper. While it remains plausible that changes in spin temperature influence the detection rate for \hi absorption at different redshifts, further work on larger samples is needed to understand the physical processes by which this might occur. 

\section{Summary and conclusions}

In this paper, we have presented the results of a search for associated \mbox{H\,{\sc i}} absorption in the GAMA 09, 12, and 15 fields, using data from the FLASH pilot survey.  

We compiled a sample of 326 targets at $z=0.42-1.0$ and detected two associated \mbox{H\,{\sc i}} absorption lines, in SDSS J090331+010847 and SDSS J113622+004852. 

Both these galaxies are massive (stellar mass $>10^{11}$M$_\odot$ at $z\sim0.5$) and contain compact, peaked-spectrum radio sources. These are relatively faint radio sources, with flux densities of 55 and 171\,mJy respectively at 855.5\,MHz (63 and 155\,mJy at 1.4 GHz). 
Both absorption lines have high optical depth, $1.77^{+0.16}_{-0.16}$ for SDSS J090331+010847 and $0.14^{+0.01}_{-0.01}$ (from 6 hours observation with SBID 13306) for SDSS J113622+004852, implying \hi column densities above $10^{21}$\,cm$^{-2}$. 

The integrated \hi optical depth of 118$\pm7$\,km\,s$^{-1}$ for SDSS J090331+010847 is the highest ever measured for an associated \hi absorption line beyond the local Universe, well  exceeding the previous highest value of $78.4\pm0.8$\,km\,s$^{-1}$ measured by \cite{Chowdhury2020} for the source J0229$+$0053 at $z=1.16$. 

We used the \prospect SED fitting code \citep{robotham2020} to model the past star-formation history and find that star-formation has been largely quenched in SDSS J090331+010847 (SFR $\sim1.47$ M$_\odot$\,yr$^{-1}$), while SDSS J113622+004852 continues to sustain a high star-formation rate (SFR $\sim69$ M$_\odot$\,yr$^{-1}$). Thus even in our current small sample we see a diversity of star-formation properties for the galaxies  detected in \hi absorption. 

By combining our data with the results from other \mbox{H\,{\sc i}} absorption searches at lower and higher redshift, we find a tendency for the detection rate for associated \hi absorption (with $\tau_{\rm pk}>0.1$) to decrease towards higher redshift. Similar results have been found in other recent studies \citep[e.g.][]{aditya2018b,Chowdhury2020}, but further work and larger samples will be needed both to establish the statistical significance of this trend and to understand the physical processes driving it. 

\section*{Acknowledgements}
RZS thanks CSIRO for their hospitality during his stay in Australia and acknowledges support from a joint SKA PhD scholarship.

RZS and MFG are supported by the National Science Foundation of China (grant 11873073), Shanghai Pilot Program for Basic Research--Chinese Academy of Science, Shanghai Branch (JCYJ-SHFY-2021-013), and the science research grants from the China Manned Space Project with NO. CMSCSST-2021-A06. This work is supported by the Original Innovation Program of the Chinese Academy of Sciences (E085021002).
RS acknowledges grant number 12073029 from the National Science Foundation of China. 

Parts of this research were supported by the Australian Research Council Centre of Excellence for All Sky Astrophysics in 3 Dimensions (ASTRO 3D), through project number CE170100013.

The Australian SKA Pathfinder is part of the Australia Telescope National Facility (grid.421683.a) which is managed by CSIRO. Operation of ASKAP is funded by the Australian Government with support from the National Collaborative Research Infrastructure Strategy. ASKAP uses the resources of the Pawsey Supercomputing Centre. Establishment of ASKAP, the Murchison Radio-astronomy Observatory and the Pawsey Supercomputing Centre are initiatives of the Australian Government, with support from the Government of Western Australia and the Science and Industry Endowment Fund. We acknowledge the Wajarri Yamatji people as the traditional owners of the Observatory site.

This research has made use of the NASA/IPAC Extragalactic Database (NED), which is operated by the Jet Propulsion Laboratory, California Institute of Technology, under contract with the National Aeronautics and Space Administration; the data products from the Wide-field Infrared Survey Explorer, which is a joint project of the University of California, Los Angeles, and the Jet Propulsion Laboratory/California Institute of Technology, funded by the National Aeronautics and Space Administration; the VizieR catalogue access tool, CDS, Strasbourg, France (DOI: 10.26093/cds/vizier);

GAMA is a joint European-Australasian project based around a spectroscopic campaign using the Anglo-Australian Telescope. The GAMA input catalogue is based on data taken from the Sloan Digital Sky Survey and the UKIRT Infrared Deep Sky Survey. Complementary imaging of the GAMA regions is being obtained by a number of independent survey programmes including GALEX MIS, VST KiDS, VISTA VIKING, WISE, Herschel-ATLAS, GMRT and ASKAP providing UV to radio coverage. GAMA is funded by the STFC (UK), the ARC (Australia), the AAO, and the participating institutions. The GAMA website is http://www.gama-survey.org/ . 

Funding for the Sloan Digital Sky Survey IV has been provided by the Alfred P. Sloan Foundation, the U.S. Department of Energy Office of Science, and the Participating Institutions. SDSS acknowledges support and resources from the Center for High-Performance Computing at the University of Utah. The SDSS web site is www.sdss.org.

SDSS is managed by the Astrophysical Research Consortium for the Participating Institutions of the SDSS Collaboration including the Brazilian Participation Group, the Carnegie Institution for Science, Carnegie Mellon University, Center for Astrophysics | Harvard \& Smithsonian (CfA), the Chilean Participation Group, the French Participation Group, Instituto de Astrofísica de Canarias, The Johns Hopkins University, Kavli Institute for the Physics and Mathematics of the Universe (IPMU) / University of Tokyo, the Korean Participation Group, Lawrence Berkeley National Laboratory, Leibniz Institut f{\"u}r Astrophysik Potsdam (AIP), Max-Planck-Institut f{\"u}r Astronomie (MPIA Heidelberg), Max-Planck-Institut f{\"u}r Astrophysik (MPA Garching), Max-Planck-Institut f{\"u}r Extraterrestrische Physik (MPE), National Astronomical Observatories of China, New Mexico State University, New York University, University of Notre Dame, Observatório Nacional / MCTI, The Ohio State University, Pennsylvania State University, Shanghai Astronomical Observatory, United Kingdom Participation Group, Universidad Nacional Autónoma de México, University of Arizona, University of Colorado Boulder, University of Oxford, University of Portsmouth, University of Utah, University of Virginia, University of Washington, University of Wisconsin, Vanderbilt University, and Yale University.

The Legacy Surveys consist of three individual and complementary projects: the Dark Energy Camera Legacy Survey (DECaLS; Proposal ID \#2014B-0404; PIs: David Schlegel and Arjun Dey), the Beijing-Arizona Sky Survey (BASS; NOAO Prop. ID \#2015A-0801; PIs: Zhou Xu and Xiaohui Fan), and the Mayall z-band Legacy Survey (MzLS; Prop. ID \#2016A-0453; PI: Arjun Dey). DECaLS, BASS and MzLS together include data obtained, respectively, at the Blanco telescope, Cerro Tololo Inter-American Observatory, NSF’s NOIRLab; the Bok telescope, Steward Observatory, University of Arizona; and the Mayall telescope, Kitt Peak National Observatory, NOIRLab. The Legacy Surveys project is honored to be permitted to conduct astronomical research on Iolkam Du’ag (Kitt Peak), a mountain with particular significance to the Tohono O’odham Nation.

NOIRLab is operated by the Association of Universities for Research in Astronomy (AURA) under a cooperative agreement with the National Science Foundation.

This project used data obtained with the Dark Energy Camera (DECam), which was constructed by the Dark Energy Survey (DES) collaboration. Funding for the DES Projects has been provided by the U.S. Department of Energy, the U.S. National Science Foundation, the Ministry of Science and Education of Spain, the Science and Technology Facilities Council of the United Kingdom, the Higher Education Funding Council for England, the National Center for Supercomputing Applications at the University of Illinois at Urbana-Champaign, the Kavli Institute of Cosmological Physics at the University of Chicago, Center for Cosmology and Astro-Particle Physics at the Ohio State University, the Mitchell Institute for Fundamental Physics and Astronomy at Texas A\&M University, Financiadora de Estudos e Projetos, Fundacao Carlos Chagas Filho de Amparo, Financiadora de Estudos e Projetos, Fundacao Carlos Chagas Filho de Amparo a Pesquisa do Estado do Rio de Janeiro, Conselho Nacional de Desenvolvimento Cientifico e Tecnologico and the Ministerio da Ciencia, Tecnologia e Inovacao, the Deutsche Forschungsgemeinschaft and the Collaborating Institutions in the Dark Energy Survey. The Collaborating Institutions are Argonne National Laboratory, the University of California at Santa Cruz, the University of Cambridge, Centro de Investigaciones Energeticas, Medioambientales y Tecnologicas-Madrid, the University of Chicago, University College London, the DES-Brazil Consortium, the University of Edinburgh, the Eidgenossische Technische Hochschule (ETH) Zurich, Fermi National Accelerator Laboratory, the University of Illinois at Urbana-Champaign, the Institut de Ciencies de l’Espai (IEEC/CSIC), the Institut de Fisica d’Altes Energies, Lawrence Berkeley National Laboratory, the Ludwig Maximilians Universitat Munchen and the associated Excellence Cluster Universe, the University of Michigan, NSF’s NOIRLab, the University of Nottingham, the Ohio State University, the University of Pennsylvania, the University of Portsmouth, SLAC National Accelerator Laboratory, Stanford University, the University of Sussex, and Texas A\&M University.

BASS is a key project of the Telescope Access Program (TAP), which has been funded by the National Astronomical Observatories of China, the Chinese Academy of Sciences (the Strategic Priority Research Program ``The Emergence of Cosmological Structures'' Grant \# XDB09000000), and the Special Fund for Astronomy from the Ministry of Finance. The BASS is also supported by the External Cooperation Program of Chinese Academy of Sciences (Grant \# 114A11KYSB20160057), and Chinese National Natural Science Foundation (Grant \# 11433005).

The Legacy Survey team makes use of data products from the Near-Earth Object Wide-field Infrared Survey Explorer (NEOWISE), which is a project of the Jet Propulsion Laboratory/California Institute of Technology. NEOWISE is funded by the National Aeronautics and Space Administration.

The Legacy Surveys imaging of the DESI footprint is supported by the Director, Office of Science, Office of High Energy Physics of the U.S. Department of Energy under Contract No. DE-AC02-05CH1123, by the National Energy Research Scientific Computing Center, a DOE Office of Science User Facility under the same contract; and by the U.S. National Science Foundation, Division of Astronomical Sciences under Contract No. AST-0950945 to NOAO.

We thank Joe Callingham and Tim Shimwell for examining LOFAR images of the field around SDSS J090331+010847 and providing the 144\,MHz upper limit plotted in Fig. \ref{fig:radio_sed_fit}. 

We thank the anonymous referee for helpful comments which have improved the paper.

\section*{Data availability}
The data underlying this article will be shared on reasonable request to the corresponding author.




\bibliographystyle{mnras}
\bibliography{gama} 

\appendix

\section{Catalogue of extended sources that have been excluded from our target sample}\label{appendix:cata_excluded}

\begin{landscape}
\begin{table} 
\centering
\caption{Radio sources with known redshifts between $z=0.42$ and 1.00, and with flux density $S_{855.5}\geq10 $ mJy that were excluded from our target sample due to their extended FIRST morphology. Column [1] lists the island IDs. Column [2] and [3] are RA and Dec. of islands. Column [4] is the component designation in the island. Column [5] and [6] are RA and Dec. of components. Column [7] is peak flux density, $S_{855.5}$. Column [8] is the cross-matched SDSS IDs.  Column [9] is redshift. Column [10] indicates whether the optical source is in the \protect\cite{ching2017} sample. Column[11] is separation between island coordinates and optical coordinates. We show here a part of the table to illustrate its content and format. For the whole table, please refer to the online version.}
\label{table:excluded-target-sources}
\begin{tabular}{lcccccccccc}
\hline \hline
Island ID &  \multicolumn{2}{c}{Island coordinates}  & Comp\_ID  &   \multicolumn{2}{c}{Component coordinates} & Flux density & SDSS ID  & Redshift & In \cite{ching2017}& Sep. \\
& RA\,[J2000] & Dec.\,[J2000] &    &RA\,[J2000]  & Dec.\,[J2000] &  mJy & & & & arcsec \\

[1]& [2] & [3] & [4] & [5] & [6] & [7] & [8] & [9]& [10]& [11]\\

\hline
SB13285\_island\_811	&08:35:33.66	&-00:07:29.35	&a	&08:35:35.37	&-00:07:42.04	&12	&J083533.58-000733.19	&0.470$^{s}$	&yes	&4.02\\
SB13285\_island\_1121	&08:36:20.94	&01:35:30.62	&a	&08:36:20.70	&01:35:22.21	&10.1	&J083621.01+013530.37	&0.640$^{s}$	&no	&1.14\\
SB13285\_island\_194	&08:38:20.99	&02:35:11.74	&a	&08:38:19.94	&02:35:23.73	&64.4	&J083820.82+023516.10	&0.428$^{g}$	&no	&5.05\\
SB13285\_island\_194	&08:38:20.99	&02:35:11.74	&b	&08:38:22.26	&02:34:57.32	&52.2	&J083820.82+023516.10	&0.428$^{g}$	&no	&5.05\\
SB13285\_island\_124	&08:38:23.22	&01:50:08.04	&a	&08:38:22.80	&01:50:18.54	&85	&J083823.10+015012.46	&0.559$^{s}$	&no	&4.74\\
SB13285\_island\_124	&08:38:23.22	&01:50:08.04	&d	&08:38:23.38	&01:50:18.71	&14.6	&J083823.10+015012.46	&0.559$^{s}$	&no	&4.74\\
SB13285\_island\_124	&08:38:23.22	&01:50:08.04	&c	&08:38:23.41	&01:50:02.10	&26.7	&J083823.10+015012.46	&0.559$^{s}$	&no	&4.74\\
SB13285\_island\_124	&08:38:23.22	&01:50:08.04	&b	&08:38:23.71	&01:49:51.58	&57.7	&J083823.10+015012.46	&0.559$^{s}$	&no	&4.74\\
SB13285\_island\_579	&08:43:25.20	&02:41:13.64	&b	&08:43:24.51	&02:41:04.09	&16.9	&J084325.19+024112.07	&0.540$^{s}$	&yes	&1.59\\
SB13285\_island\_579	&08:43:25.20	&02:41:13.64	&a	&08:43:25.80	&02:41:21.92	&22.2	&J084325.19+024112.07	&0.540$^{s}$	&yes	&1.59\\
SB13285\_island\_480	&08:44:26.77	&02:27:57.03	&b	&08:44:26.19	&02:28:06.79	&17.7	&J084426.64+022757.28	&0.450$^{s}$	&no	&2.09\\
SB13285\_island\_480	&08:44:26.77	&02:27:57.03	&a	&08:44:27.25	&02:27:51.80	&18	&J084426.64+022757.28	&0.450$^{s}$	&no	&2.09\\
SB13285\_island\_763	&08:44:34.92	&03:10:20.45	&a	&08:44:34.89	&03:10:20.49	&16.6	&J084434.86+031020.36	&0.723$^{s}$	&no	&0.87\\
SB13285\_island\_96	&08:48:13.79	&01:15:11.40	&b	&08:48:13.12	&01:14:11.34	&56.8	&J084813.71+011502.36	&0.570$^{c}$	&yes	&9.12\\
SB13285\_island\_96	&08:48:13.79	&01:15:11.40	&d	&08:48:13.39	&01:14:34.35	&36	&J084813.71+011502.36	&0.570$^{c}$	&yes	&9.12\\
SB13285\_island\_96	&08:48:13.79	&01:15:11.40	&e	&08:48:13.98	&01:15:34.50	&25.6	&J084813.71+011502.36	&0.570$^{c}$	&yes	&9.12\\
SB13285\_island\_96	&08:48:13.79	&01:15:11.40	&c	&08:48:14.29	&01:16:00.44	&45.5	&J084813.71+011502.36	&0.570$^{c}$	&yes	&9.12\\
SB13285\_island\_96	&08:48:13.79	&01:15:11.40	&a	&08:48:14.47	&01:16:06.24	&83.6	&J084813.71+011502.36	&0.570$^{c}$	&yes	&9.12\\
SB13285\_island\_604	&08:48:27.03	&00:01:40.67	&b	&08:48:26.44	&00:01:36.42	&18.1	&J084826.98+000141.36	&0.569$^{s}$	&yes	&1.01\\
SB13285\_island\_604	&08:48:27.03	&00:01:40.67	&a	&08:48:27.53	&00:01:43.95	&20.7	&J084826.98+000141.36	&0.569$^{s}$	&yes	&1.01\\
\multicolumn{11}{c}{......}\\
\hline\end{tabular}
\\
\medskip
$^{s}$redshift from SDSS.
$^{g}$redshift from GAMA.
$^{n}$redshift from NED. 
$^{c}$ redshift from \cite{ching2017}.
\end{table}
\end{landscape}


\bsp	
\label{lastpage}
\end{document}